\documentclass[twocolumn,showpacs,preprintnumbers,amsmath,amssymb]{revtex4}
\usepackage{epsf}
\usepackage{graphicx}
\usepackage{dcolumn}
\usepackage{bbm}

\newcommand{\bi}{\bigskip}
\newcommand{\no}{\noindent}
\newcommand{\be}{\begin{eqnarray}}
\newcommand{\ee}{\end{eqnarray}}
\newcommand{\hk}{\hspace{0.1cm}}
\newcommand{\hs}{\hspace{0.5cm}}

\newcommand{\rk}{\right)}
\newcommand{\lk}{\left(}

\newcommand{\sli}{\sum\limits}

\newcommand{\il}{\int\limits}

\def\R{{\mathbb{R}}}

\begin{document}

\title{Variational solution of the Yang-Mills Schr\"odinger equation in Coulomb gauge\footnote{Supported by DFG-Re 856}}
\date{\today}

\author{C.~Feuchter and H.~Reinhardt}
\affiliation{Institut f\"ur Theoretische Physik\\
Auf der Morgenstelle 14\\
D-72076 T\"ubingen\\
Germany}




\bi

\no
\begin{abstract}
The Yang-Mills Schr\"odinger equation is solved in Coulomb gauge for the vacuum
by the variational principle using an ansatz for the wave functional, which is
strongly peaked at the Gribov horizon. A coupled set of Schwinger-Dyson
equations for the gluon and ghost propagators in the Yang-Mills vacuum as well as
for the curvature of gauge orbit space is derived and solved in one-loop
approximation. We find an infrared suppressed gluon propagator, an 
infrared singular ghost propagator and a almost linearly rising confinement potential.
\end{abstract}
\bi

\no
\maketitle
\bi

\no
\section{Introduction \label{sec1}}
\bi

\no
To understand the low energy sector of QCD is one of the most challenging
problems in quantum field theory. Nowadays, quantum field theory and, in
particular, QCD is usually studied within the functional integral approach. This
approach is advantagous for a perturbative calculation, where it leads
automatically to a Feynman diagrammatic expansion. Within this approach
asymptotic freedom of QCD was shown \cite{R1}, which manisfests itself in deep
inelastic scattering experiments. In addition, the functional integral
 approach is
the basis for numerical lattice calculations \cite{R2}, the only rigorous
non-perturbative approach available at the moment. 
These lattice methods have provided considerable 
insights into the nature of the Yang-Mills vacuum. Lattice
investigations performed over the last decade, have accumulated evidence for two
confinement scenarios: the dual Meissner effect based on a condensate of
magnetic monopoles \cite{R3} 
and the vortex condensation picture \cite{R4} (For a recent review see ref.
\cite{R5}). In both cases the
Yang-Mills functional integral is dominated in the infrared sector by
topological field configurations (magnetic monopoles \cite{RX1} 
or center vortices \cite{RX2}), which
seem to account for the string tension, i.e. for the confining force. Yet
another confinement mechanism was proposed by Gribov \cite{R22}, further elaborated
by Zwanziger \cite {R23} and tested in lattice calculations \cite{R24}. 
This mechanism is based on the infrared dominance of the field
configurations near the Gribov horizon in Coulomb gauge. This mechanism of
confinement is compatible with the center vortex and magnetic pictures, 
given the fact, that lattice center vortex and magnetic monopole 
configurations lie on the Gribov horizon \cite{R26}. \\
%
Despite the
great successes of lattice calculations in the exploration of strong interaction
physics \cite{R6},
a complete
understanding of the Yang-Mills theory will probably not be provided by the
lattice simulations alone, but requires also analytical tools. Despite of its
success in quantum field theory, in particular in perturbation theory and its
appeal in semi-classical and topological considerations \cite{R9}, 
the path integral
approach may not be the most economic method for analytic studies of
non-perturbative physics. As an example consider the hydrogen atom.
Calculating its electron spectrum exactly in the path integral approach is
exceedingly complicated \cite{R7}, 
while the exact spectrum can be obtained easily by
solving the Schr\"odinger equation. 
One might therefore wonder, whether the Schr\"odinger
equation is also the appropriate tool to study the low-energy sector of
Yang-Mills theory, and in particular of QCD. \\
%
The Yang-Mills Schr\"odinger equation is based on the canonical quantization in
Weyl gauge $A_0 = 0$ \cite{R8}, where Gau\ss' law has to be enforced as a
constraint to the wave functional to guarantee gauge invariance. The
implementation of Gau\ss' law is crucial. This is because any violation of Gau\ss'
law generates spurious color charges during the time evolution. These spurious
color charges can screen the actual color charges and thereby spoil 
confinement \footnote{One of the authors (H.R.) is
indepted to the late Ken Johnson for elucidating discussions on 
this subject.}. Several approaches have
been advocated to explicitly resolve Gau\ss' law by changing variables resulting
in a description in terms of a reduced number of unconstrained variables. This
can be accomplished either by choosing a priori gauge invariant variables
\cite{R10} or by fixing the gauge, for example, to unitary gauge \cite{R11},
 to Coulomb gauge \cite{R12} 
 or to a modified version of axial gauge \cite{R13}.
In particular, the Yang-Mills Hamiltonian resulting after eliminating the gauge
degrees of freedom in Coulomb gauge, was derived in ref. \cite{R12}.
Alternatively, one has attempted to project the Yang-Mills wave functional onto gauge invariant
states (which a priori fulfill Gau\ss' law) \cite{R14}. The equivalence between
gauge fixing and projection onto gauge invariant states can be seen by noticing
that the Faddeev-Popov determinant provides the Haar measure of the gauge group
\cite{R15}. \\
%
In this paper we will variationally solve the stationary Yang-Mills Schr\"odinger equation 
in Coulomb gauge for the vacuum. Such a variational approach was previously 
studied in refs. \cite{R16}, \cite{R17}, 
where a Gaussian ansatz for the Yang-Mills wave functional was used.
Our approach is conceptually simular to,
but differs essentially from refs. \cite{R16,R17} in two respects: 
i) we use a different ansatz for the trial wave functional and 
ii) we include fully the curvature of the space of gauge orbits induced by
the Faddeev-Popov determinant.
We use a vacuum wave functional, which is strongly peaked at the
Gribov horizon. Such a wave functional is motivated by the results of ref.
\cite{R19} 
and by the fact, that the dominant infrared degrees of freedom like center
vortices lie on the Gribov horizon \cite{R26}. 
The Faddeev-Popov determinant was completely ignored in ref. \cite{R16} and only partly included
in ref. \cite{R17}. (In addition, ref. \cite{R17} uses the angular approximation).
We will find however, that a full inclusion of the curvature induced by the Faddeev-Popov determinant
is absolutely crucial and vital for the infrared regime and, in particular, for the confinement 
property of the Yang-Mills theory.
The organization of the paper is as follows: \\

In section \ref{sec2} we briefly review the hamiltonian formulation of Yang-Mills theory
in Coulomb gauge and fix our notation. In section \ref{sec3} we present our vacuum wave
functional and calculate the relevant expectation values. 
The vacuum energy functional is calculated and minimized in section \ref{sec4}
resulting in a set of coupled Schwinger-Dyson equations for the gluon energy,
the ghost and Coulomb form factors and the curvature in gauge orbit space. 
The asymptotic behaviours
of these quantities in both the ultraviolet and infrared regimes are 
investigated in section \ref{sec5}. Section \ref{sec6} is devoted to the renormalization of the
Schwinger-Dyson equations. The numerical solutions to the renormalized
Schwinger-Dyson equations are presented in section \ref{sec7}. Finally in section \ref{sec8} we
present our results for the static Coulomb potential.
Our conclusions are given in section \ref{sec9}.
A short summary of our results has been previously reported in ref. \cite{R9a}.
\bi

\no
\section{Hamiltonian formulation of Yang-Mills theory in Coulomb gauge \label{sec2}}
\bi

\no
The canonical quantization of gauge theory is usually performed in the Weyl
gauge $A_0 = 0$. In this gauge the spatial components of the gauge field
${\bf{A}} ({\bf x})$ are the ``cartesian'' 
coordinates and the corresponding canonically
conjugated momenta
\be
\label{1}
\Pi^a_k ({\bf x}) = \frac{\delta}{i \delta A^a_k ({\bf x})} \hk ,
\ee
defined by the equal time commutation relation
\be
\label{2}
\left[ A^a_k ({{\bf x}}), \Pi^b_l ({\bf y}) \right] = 
i \delta^{a b} \delta_{k l}
\delta \lk {\bf x} - {\bf y} \rk
\ee
represent the color electric field. 
The Yang-Mills Hamiltonian is then given by
\be
\label{3} 
H = \frac{1}{2} \int d^3 x \left[ \Pi^a_k ({\bf x})^2 +
B^a_k ({\bf x})^2 \right] \hk ,
\ee
where
\be
\label{4}
B_k = \frac{1}{2} \epsilon_{k i j} F_{i j} \hk , \hk  g F_{i j} =  
\left[ D_i, D_j
\right] \hk , \hk D_i = \partial_i + g A_i
\ee
is the colormagnetic field with $F_{i j}$ being the non-Abelian field strength
and $D_i$ the covariant derivative. We use anti-hermitian 
 generators $T^a$ of the gauge group with normalization $ tr(T^a T^b) = - \frac{1}{2} \delta^{a b}$
 $(A = A^a T^a)$. \\
%
The Hamiltonian
(\ref{3}) is invariant under spatial gauge transformations $U ({\bf x})$
\be
\label{5}
A \to A^U = U A U^\dagger + \frac{1}{g} U \partial U^\dagger \hk .
\ee
Accordingly the Yang-Mills wave functional $\Psi [A] = \langle A | \Psi \rangle$
 can change
only by a phase, which is given by
\be
\label{6}
\Psi [A^U] = e^{i \Theta n [U]} \Psi [A] \hk ,
\ee
where $\Theta$ is the vacuum angle and $n [U]$ denotes the winding number of the
mapping $U ({\bf x})$ from the 3-dimensional space $\R^3$ 
(compactified to $S_3$) into
the gauge group. In this paper we will not be concerned with topological aspects
of gauge theory and confine ourselves to ``small'' gauge transformations with $n
[U]= 0$. \\
%
%
Invariance of the wave functional under ``small'' gauge
transformations is guaranteed by Gau\ss' law
\be
\label{7}
\hat{D}_k(A) \Pi_k \Psi [A] = \rho_m \hk ,
\ee
where $\rho_m$ is the color density of the matter (quark) field and 
$\hat{D}_i(A)$ denotes the covariant derivative in the adjoint representation
of the gauge group 
\be
\label{10}
\hat{D}_i(A) = \partial_i + g \hat{A}_i \hk , \hk \hat{A}_i = 
A^{a}_i \hat{T}^a \hk ,
\hk ( \hat{T}^a )^{b c} = f^{b a c} 
\ee
with $f^{a b c}$ being the structure constant of the gauge group. Throughout
the paper we use a hat ``\^{}'' to denote quantities defined in the adjoint
representation of the gauge group.
The Gau\ss' law constraint (\ref{7}) 
on the Yang-Mills wave functional can be resolved by fixing the
residual gauge invariance, eq. (\ref{5}). This eliminates the unphysical gauge
degrees of freedom and amounts to a change of coordinates from the ``cartesian''
coordinates to curvilinear ones, which introduces a non-trivial Jacobian
(Faddeev-Popov determinant) ${\cal{J}} [A]$. In this paper we shall use the Coulomb
gauge
\be
\label{8}
\partial_k A_k ({\bf x}) = 0 \hk .
\ee
In this gauge the physical degrees of freedom are the transversal components of
the gauge fields
\be
\label{8a}
A^\perp_i ({{\bf x}}) = t_{i j} ({{\bf x}}) A_j ({{\bf x}}) \hk ,
\ee
where
\be
\label{8b}
t_{i j} ({\bf x}) = \delta_{i j} - \frac{\partial_i \partial_j}{\partial^2}
\ee
\vspace{0.2cm}
is the transversal projector. The corresponding canonical momentum is given by
\be
\label{8c}
\Pi^{\perp a}_i ({\bf x}) = t_{i k} ({\bf x}) \frac{\delta}{i \delta A^a_k ({\bf x})} :=
\frac{\delta}{i \delta A^{\perp a}_i ({\bf x})}
\ee
and satisfies the equal-time commutation relation
\be
\label{8d}
\left[ A^{\perp a}_i ({\bf x}) \hk , \hk \Pi^{\perp b}_j ({\bf x}') \right]  = i 
\delta^{a b} t_{i j} ({\bf x}) \delta
\lk {{\bf x} - {\bf x}'} \rk\hk 
\ee
In Coulomb gauge (\ref{8}) the Yang-Mills Hamiltonian is given by 
\cite{R12}
\begin{widetext}
\be
\label{11}
H & = & \frac{1}{2} \int d^3 x \left[ {\cal{J}}^{- 1} [A^\perp] \Pi^a_i ({\bf x}) 
{\cal{J}} [A^\perp] \Pi^a_i
({\bf x})  +  B^a_i ({{\bf x}})^2 \right] \nonumber\\
& & + \frac{g^2}{2} \int d^3 x \int d^3 x' {\cal{J}}^{- 1} [A^\perp] \rho^a ({\bf x}) 
F^{a b} \lk {{\bf x}},
{{\bf x}'} \rk {\cal{J}} [A^\perp] \rho^b ({{\bf x}'}) \hk .
\ee
\end{widetext}
where
\be
\label{9}
{\cal{J}} [A^\perp] = Det \lk - \partial_i \hat{D}_i \rk \hk 
\ee
is the Faddeev-Popov determinant with 
$\hat{D_k} \equiv \hat{D_k}( A^{\perp} )$ .

The kinetic part of the hamiltonian (first term) is reminiscent to the functional version of the
Laplace-Beltrami operator in curvilinear space. The magnetic energy (second
term) represents the potential for the gauge field. Finally, the last term
arises by expressing the kinetic part of eq. (\ref{3}) in terms of the transversal gauge
potentials (curvilinear coordinates) 
satisfying the Coulomb gauge, i.e. by resolving Gau\ss'  law \cite{R12}. 
This term describes the interaction of
static non-Abelian color (electric) charges with density
\be
\label{12}
\rho^a ({{\bf x}}) = - \hat{A}^{\perp a b}_i ({{\bf x}}) \Pi^b_i ({{\bf x}}) + \rho^a_m ({\bf x})
\ee
through the Coulomb propagator 
\be
\label{13}
F^{a b} \lk {{\bf x}, {\bf x}'} \rk = \langle
 {{\bf x}} a | \lk - \hat{D}_i 
\partial_i
\rk^{- 1} ( - \partial^2) \lk - \hat{D}_j  \partial_j
\rk^{- 1} | {{\bf x}'} b \rangle \hk \hk
\ee
which is the non-Abelian counter part of the usual Coulomb propagator $\langle
 {{\bf x}} |
\frac{1}{- \partial^2} | {{\bf x}'} \rangle$ 
in QED and reduces to the latter for the
perturbative vacuum $A^\perp = 0$. If not stated otherwise, in the following we will assume
$\rho_m = 0$. \\

Since the Hamiltonian (\ref{11}) does not change when ${\cal{J}} [A^\perp]$ 
is multiplied by
a constant we can rescale the Jacobian ${\cal{J}}$ by the irrelevant constant $Det \lk -
\partial^2 \rk$
\be
\label{x501}
{\cal{J}} [A^\perp] = 
\frac{Det \lk - \hat{D}_i \partial_i \rk}{Det \lk - {\partial^2}
 \rk}
\ee
such that ${\cal{J}} [A^\perp = 0] = 1$. \\
The Faddeev-Popov matrix $\lk - \hat{D}_i \partial_i
\rk$ represents the metric tensor in the color 
space of gauge connections satisfying
the Coulomb gauge. Accordingly its determinant enters the measure of the integration 
over the space of transversal gauge connections 
and the matrix element of an observable $O [A^\perp,\Pi]$
between wave functionals $\Psi_1 [A^\perp] , \Psi_2 [A^\perp]$ is defined by
\begin{align}
\label{14}
\langle \Psi_1 | O | \Psi_2 \rangle 
= \int {\cal D} A^{\perp} {\cal{J}} [A^{\perp}] \Psi^*_1 
[A^{\perp}]
O [A^{\perp},\Pi] \Psi_2 [A^{\perp}] \hk .
\end{align}
Here, the integration is over transversal gauge potentials $A^{\perp}$ 
 only. The same expression (\ref{14}) is obtained by starting from the matrix
 element in Weyl gauge (with gauge invariant wave functionals) and implementing
 the Coulomb gauge by means of the Faddeev-Popov method. \\
 
Like in the treatment of spherically symmetric systems in quantum mechanics 
it is
convenient to introduce ``radial'' wave functions by
\be
\label{15}
\tilde{\Psi} [A^\perp] = {\cal{J}}^{\frac{1}{2}} [A^\perp] \Psi [A^\perp]
\ee
and transform accordingly the observables
\be
\label{16}
\tilde{O} = {\cal{J}}^{\frac{1}{2}} [A^\perp]  O {\cal{J}}^{- \frac{1}{2}} [A^\perp] \hk .
\ee
Formally, 
 the Jacobian then disappears from the matrix element
\be
\label{17}
\langle \Psi_1 | O | \Psi_2 \rangle
 =  \int  D A^\perp \tilde{\Psi}_1^* [A^\perp] \hk \tilde{O}  \hk
 \tilde{\Psi}_2 [A^\perp] : =  \langle
  \tilde{\Psi}_1 | \tilde{O} | \tilde{\Psi}_2 \rangle 
 \hk . \hk \hk
\ee
%
In the present paper we are interested in the vacuum structure of Yang-Mills
theory. The vacuum wave functional, defined as solution to the Yang-Mills
Schr\"odinger equation
\be
\label{18}
H \Psi = E \Psi
\ee
for the lowest energy eigenstate, can be obtained from the variational principle
\be
\label{19}
\frac{\langle \Psi | H | \Psi \rangle}{\langle \Psi | \Psi \rangle } \longrightarrow \mbox{min} \hk .
\ee
%
To work out the expectation value of the Hamiltonian $\tilde{H} =
{\cal{J}}^{\frac{1}{2}} H {\cal{J}}^{- \frac{1}{2}}$
it is convenient to perform a partial (functional) integration in the kinetic
and Coulomb terms. Assuming as usual that the emerging surface terms vanish, one
finds
\be
\label{20}
\langle \Psi | H | \Psi \rangle = 
\langle \tilde{\Psi} | \tilde{H} | \tilde{\Psi} \rangle = E_k + E_p + E_c \hk ,
\ee
where
\begin{widetext}
\begin{align}
\label{21}
E_k &= \frac{1}{2} \int {\cal D} 
A^\perp  \int d^3 x \left[ \tilde{\Pi}^a_i ({\bf x}) \tilde{\Psi} [A^\perp] \right]^* \left[ 
\tilde{\Pi}^a_i ({\bf x}) \tilde{\Psi} [A^\perp] \right] \\
\label{22}
E_p &= \frac{1}{2 } \int {\cal D} A^\perp \int d^3 x \hk \tilde{\Psi}^* 
[A^\perp] B^a_i
({\bf x})^2 \tilde{\Psi} [A^\perp] 
\\
\label{23}
E_c &= - \frac{g^2}{2} \int {\cal D} A^\perp \int d^3 x \int d^3 x' \left[
\tilde{\Pi}^c_i ({\bf x}) \tilde{\Psi} [A^\perp] \right]^*
\hat{A}^{\perp c a}_{i} ({\bf x}) F^{a b} ({\bf x}, {\bf x}') 
\hat{A}^{\perp b d}_j 
({\bf x}') \left[
\tilde{\Pi}^d_j ({\bf x}') \tilde{\Psi} [A^\perp] \right] \hk ,
\end{align}
\end{widetext}
where $\tilde{\Pi}$ is defined below in eq. (\ref{24}).
Note, integration is over the transversal part $A^{\perp}$ only. In order to
prevent the equations from getting cluttered we will in the following often
write $A$ instead of $A^\perp$, and $\Pi$ instead of $\Pi^\perp$, 
but it will be always clear from the context,
when the transversal part is meant. 

The transformed momentum operator $\tilde{\Pi}^a_i (x)$ is
explicitly given by
\begin{align}
\label{24}
\tilde{\Pi}^a_k ({\bf x}) & = {\cal{J}}^{\frac{1}{2}} [A^\perp] \Pi^{\perp a}_k ({\bf x}) 
{\cal{J}}^{- \frac{1}{2}} [A^\perp]
\nonumber\\
& = \Pi^{\perp a}_k ({\bf x}) - \frac{1}{2} \Pi^{\perp a}_k ({\bf x}) \ln {\cal{J}} [A^\perp] \nonumber\\
& = \Pi^{\perp a}_k ({\bf x})  + \frac{g}{2 i} t_{k l} ({\bf x}) \hk
tr \left[ \hat{T}^a \lk \partial^{{\bf y}}_l G ({\bf y}, {\bf x}) \rk_{{\bf y}
= {\bf x}} \right] \hk ,
\end{align}
where we have introduced the inverse of the Faddeev-Popov operator
\be
\label{25}
G  =  \lk - \hat{D}_i \partial_i 
\rk^{- 1}  \hk ,
\ee
which is a matrix in color and coordinate space
\be
\nonumber
\langle {\bf x} a | G | {\bf x}' b \rangle = 
\langle {\bf x} a | \lk - \hat{D}_i \partial_i
\rk^{- 1} | {\bf x}' b \rangle := G^{a b} \lk {\bf x},{\bf x}' \rk  \hk .
\ee
Its vacuum expectation value (to be defined below) represents the ghost
propagator. Expanding this quantity in terms of the gauge field, we obtain
\be
\label{26}
G  =  \lk - {\partial^2} - g \hat{A}_i  \partial_i \rk^{- 1}
 =  G_0 \sli^\infty_{n = 0} \lk g \hat{A}_i \partial_i G_0 \rk^n
\hk ,
\ee
where
\be
\label{27}
G_0 \lk {\bf x} , {\bf x}' \rk = \langle {\bf x} | \lk - {\partial^2} \rk^{- 1} |
{\bf x}' \rangle = \frac{1}{4 \pi | {\bf x} - {\bf x}'|}
\ee
is the free ghost propagator, which is nothing but the ordinary static Coulomb
propagator in QED. \\
%
From the expansion, eq. (\ref{26}), it is seen, that the
ghost propagator satisfies the identity
\be
\label{55}
G = G_0 + G_0 g \hat{A}_i \partial_i G \hk ,
\ee
where we have used the usual short hand matrix notation in functional and color 
space.

\bi

\no
\section{The vacuum wave functional and propagators \label{sec3}}
\bi

\no
So far all our considerations have been exact, i.e. no approximation has been
introduced. To proceed further, we have to specify the (vacuum) wave functional
$\tilde{\Psi} [A]$. 
\bi

\no
\subsection{The vacuum wave functional \label{sec3a}}
\bi

\no
Inspired by the known exact wave functional of QED in
Coulomb gauge \cite{R18}, 
we will consider a Gaussian ansatz in the transversal
gauge fields
\begin{widetext}
\be
\label{46x}
\tilde{\Psi} [A] = \langle A | \omega \rangle = {\cal N} \exp \left[ - \frac{1}{2} \int d^3
x \int d^3 x' A^{\perp a}_i  ({\bf x}) \omega
 ({\bf x}, {\bf x}') A^{\perp a}_i  ({\bf x}')
\right] \hk ,
\ee
\end{widetext}
where $ {\cal N}$ is a normalization constant chosen so that $\langle \tilde{\Psi} |
\tilde{\Psi} \rangle = 1$. By translational and rotational 
invariance the integral kernel in the
exponent of the Gaussian wave functional $\omega 
({\bf x}, {\bf x}')$ can depend only
on $|{\bf x} - {\bf x}'|$. For simplicity we have chosen the integral kernel
$\omega 
({\bf x}, {\bf x}')$ to be a color and Lorentz scalar. This is justified by 
isotropy of color and Lorentz space. \\
%
Of course, such a Gaussian ansatz is ad hoc at this stage and can be justified
only a posteriori. Let us also stress, that we are 
using the Gaussian ansatz, eq.
(\ref{46x}), for the radial wave function $\tilde{\Psi} [A]$, eq. (\ref{15}),
which is normalized with a ``flat'' integration measure, see eq. (\ref{17}).
The original wave function
\be
\label{47x}
\Psi [A] = {\cal{J}}^{- \frac{1}{2}} [A] \tilde{\Psi} [A]
\ee
contains besides the Gaussian $\tilde{\Psi} [A]$ (\ref{46x})
an infinite power series in the gauge
potential $A$. Furthermore, since the Jacobian ${\cal{J}} [A]$ (Faddeev-Popov
determinant) vanishes at the Gribov horizon, our wave functional is strongly
peaked at the Gribov horizon. It is well known, that the infrared dominant
configurations come precisely from the Gribov horizon \cite{R19}. In addition,
the center vortices, which are believed to be the ``confiner'' in the Yang-Mills
vacuum \cite{R4} all live  on the Gribov horizon \cite{R26}. Furthermore, the
wave functional (\ref{47x}), being divergent on the Gribov horizon, identifies
all gauge configurations on the Gribov horizon, in particular those which are
gauge copies of the same orbit. This identification is absolutely necessary to
preserve gauge invariance. (The identificaton of all gauge configurations on the
Gribov horizon also topologically compactifies the first Gribov region.) 
Therefore we
prefer to make the Gaussian ansatz for the radial wave function $\tilde{\Psi}
[A]$, instead for $\Psi [A]$. Our wave functional thus drastically differs from
the one used in refs. \cite{R16}, \cite{R17}, where
an Gaussian ansatz was used for $\Psi [A]$. \\

We should also mention that the wave functional although being divergent at the
Gribov horizon it is obviously normalizable. In principle, a wave functional being
peaked at the Gribov horizon can of course have a more general form than the one given
by eq. (\ref{47x}). A somewhat more general wave functional 
\be
\label{47xy}
\Psi [A] = {\cal{J}}^{- \alpha} [A] \tilde{\Psi} [A]
\ee
would leave the power $\alpha$ as a variational parameter, which is then determined by 
minimizing the vacuum energy (density).
This would lead to a more optimized wave functional, which for $\alpha > 0$ still expresses the
dominance of the field configurations on the Gribov horizon. Such investigations are under
way. In the present paper we restrict however ourselves to $\alpha = \frac {1}{2}$, which 
simplifies the calculations a lot. 

We will use the Gaussian ansatz (\ref{46x}) 
as a trial wave function for the
Yang-Mills vacuum and determine the integral kernel $\omega ({\bf x}, {\bf x}')$ from the
variational principle, minimizing the vacuum energy density. 
The use of the Gaussian wave functional makes the
calculation feasible in the sense that Wick's theorem holds: the expectation value
of an ensemble of field operators can be expressed by the free (static) 
gluon propagator
\be
\label{5-1}
\langle A^{\perp a}_i ({\bf x}) A^{\perp b}_j ({\bf x}') \rangle_\omega &:=&  \langle
\omega | A^{\perp a}_i ({\bf x}) A^{\perp b}_j ({\bf x}') | \omega \rangle  \nonumber\\
&=& 
\frac{1}{2} \delta^{a b} t_{i j} ({\bf x}) \omega^{- 1} 
({\bf x}, {\bf x}') \hk .
\ee
The expectation value of an odd number of field operators obviously vanishes for 
the Gaussian wave functional. \\
%
To facilitate the evaluation of expectation values of field operators, we
introduce the generating functional
\begin{widetext}
\be
\label{5-7}
Z [j] & = & \langle \tilde{\Psi} | \exp \left[ { \int d^3 x j^a_i ({\bf x}) A^{\perp a}_i 
({\bf x})} \right] | \tilde{\Psi} \rangle
\nonumber\\
& = & {\cal N}^2 \int D A^{\perp} \exp \left[ - \int d^3 x \int
d^3 x' 
 A^{\perp a }_i ({\bf x} ) \omega 
({\bf x}, {\bf x}') A^{\perp a}_i ({\bf x}') +  \int
d^3 x j^a_i ({\bf x}) A^{\perp a}_i ({\bf x}) \right] \hk 
\ee
\end{widetext}
where the normalization constant ${\cal N}$ guarantees that $Z [j = 0] = 1$.
Carrying out the Gaussian integral one obtains
\begin{widetext}
\begin{align}
\label{5-8}
Z [j] = \exp \left[\frac{1}{4} \int d^3 x \int
d^3 x' j^a_i  ({\bf x}) t_{i j} ({\bf x}) 
\omega^{- 1} ( {\bf x}, {\bf x}' )
j^a_j ({\bf x}' ) \right] \hk.
\end{align}
\end{widetext}
The expectation value of any functional of the gauge field, $O [A]$,
 is then given
by
\be
\label{5-9}
\langle O [A] \rangle_\omega & = & \langle \omega | O [A] | \omega \rangle  = 
 \lk O \left[\frac{\delta}{\delta j} \right] Z [j] \rk_{j = 0} \hk .
\ee
This is basically the functional form of Wick's theorem. An alternative form of
this theorem, which will be usefull in the following, can be obtained by using
the identity
\be
\label{5-10}
F  \lk \frac{\partial}{\partial x} \rk G (x) = \left[ G \lk
\frac{\partial}{\partial y} \rk F (y) e^{x y} \right]_{y = 0} \hk .
\ee
which can be proved by Fourier transformation. Applying this to eq. (\ref{5-9}) we obtain
\begin{widetext}
\be
\label{5-11}
\langle O [A] \rangle_\omega = \left\{ \exp
\left[ \frac{1}{4} {\int d^3 x \int d^3 x' \frac{\delta}{\delta
A^{\perp a}_i ({\bf x})} t_{i j} ({\bf x}) \omega^{- 1} \lk {\bf x}, {\bf x}' \rk
\frac{\delta}{\delta A^{\perp a}_j ({\bf x}')}} \right] O [A] \right\}_{A = 0} \hk .
\ee
\end{widetext}
%
Since the kernel $\omega 
({\bf x}, {\bf x}')$ 
depends only on $|{\bf x} - {\bf x}' |$, it is convenient to
go to momentum space by Fourier transformation
\be
\label{5-2}
A^a({\bf x}) = \int \frac{d^3 k}{(2 \pi)^3} e^{i {\bf k} {\bf x}} A^a ({\bf k})
 \hk .
\ee
In momentum space the wave functional (\ref{46x}) reads
\be
\label{5-3}
\langle A | \omega \rangle = {\cal N} \exp \left[ - \frac{1}{2} 
\int \frac{d^3 k}{(2 \pi)^3}
A^a_i ({\bf k}) t_{i j} ({\bf k}) \omega ({\bf k}) A^a_j (- {\bf k}) \right] \hk 
\ee
and the free gluon propagator (\ref{5-1}) becomes
\be
\label{5-4}
\langle \omega | A^a_ i ({\bf k}) A_j (- {\bf k}') | \omega 
\rangle = \delta^{a b} \frac{t_{i j}
({\bf k})}{2 \omega ({\bf k})} (2 \pi)^3 \delta ({\bf k} - {\bf k}') \hk ,
\ee
where $\omega ({\bf k})$ represents the energy of a gluon 
with momentum ${\bf k}$.
In order that our trial wave function,  eq. (\ref{46x}), 
is integrable $\omega ({\bf k})$
has to be strictly positive definite. 
\bi

\no
\subsection{The ghost propagator and the ghost-gluon vertex \label{sec3b}}
\bi

\begin{figure}
\includegraphics [scale=0.25] {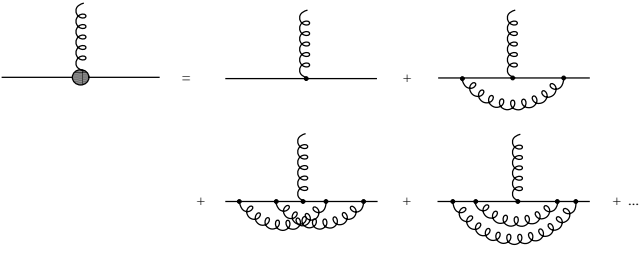}
\caption{ \label{68fig} Diagrammatic expansion of the ghost-gluon vertex.
Throughout the paper full and curly lines stand, respectively, for the full ghost and gluon 
propagators. Furthermore, dots and fat dots represent, repectively, bare and full ghost-gluon vertices.
}
\end{figure}

\no
For later use, let us consider the vacuum expectation value of the inverse 
Faddeev-Popov operator (\ref{25})
\be
\label{5-5}
 \langle \omega | G | \omega \rangle = \langle G \rangle_\omega := G_\omega \hk ,
\ee
which we refer to as ghost propagator, although we will not explicitly introduce ghost
fields. They are not needed in the present operator approach. In the functional
integral in Coulomb gauge, eq. (\ref{5-5}) would enter as the propagator 
of the ghost field.
Taking the expectation value of eq. (\ref{55}) and using the fact, that the free
ghost propagator $G_0$ does not depend on the dynamical gauge field, so that
$G_0 | \omega \rangle = | \omega \rangle G_0$, we obtain
\be
\label{56}
G_\omega = G_0 + G_0 \left\langle g \hat{A}_i \partial_i G \right\rangle_\omega \hk .
\ee
The latter expectation value can be worked out, in principle, by inserting the
expansion (\ref{26}) and using Wick's theorem. The result can be put into a
compact form by defining the one particle irreducible ghost self-energy $\Sigma$
by
\be
\label{57}
\left\langle g \hat{A}_i \partial_i G \right\rangle_\omega = \Sigma \langle G \rangle_\omega = \Sigma G_\omega
\hk .
\ee
It is then seen that $G_\omega$ satisfies the usual Dyson equation
\be
\label{58}
G_\omega = G_0 + G_0 \Sigma G_\omega \hk .
\ee
We will later also need the functional derivative of the ghost propagator
\begin{widetext}
\be
\label{y1}
&  &
\left\langle \frac{\delta}{\delta A^a_k ({\bf x})} G \lk {{\bf x}}_1, {{\bf x}}_2 
\rk \right\rangle_\omega 
 =  - \left\langle \int d^3 y_1 \int d^3 y_2 G \lk {{\bf x}}_1, {{\bf y}}_1 \rk  
\frac{\delta G^{- 1} \lk {\bf y}_1,
{\bf y}_2 \rk}{\delta A^a_k ({{\bf x}})} G \lk {\bf y}_2, {{\bf x}}_2 
\rk \right\rangle_\omega \hk .
\ee
\end{widetext}
This expectation value 
can in principle be evaluated by using 
the expansion (\ref{58}) for the ghost propagator 
and applying Wick's theorem. The emerging Feynman
diagrams have the generic structure illustrated in figure \ref{68fig}. They describe the
propagation of the ghost followed by an interaction with an external gluon and
subsequent propagation of the ghost. 
%
We therefore define the ghost-gluon vertex $\Gamma^a_k ({\bf x})$ by
\begin{widetext}
\be
\label{y2}
\left\langle \frac{\delta}{\delta A^a_k ({{\bf x}})} G \lk {{\bf x}}_1, {{\bf x}}_2 \rk 
\right\rangle_\omega = \int d^3 y_1 \int d^3 y_2 
G_\omega \lk {{\bf x}}_1, {{\bf y}}_1 \rk   \Gamma^a_k \lk {\bf y}_1, {\bf y}_2; 
{{\bf x}} \rk G_\omega \lk {\bf y}_2, {{\bf x}}_2
\rk \hk .
\ee
\end{widetext}
The ghost-gluon vertex $\Gamma^a_k ({\bf x})$ is given by the series of diagrams shown
in figure \ref{68fig}. Comparison of eqs. (\ref{y1}) and (\ref{y2}) shows, that the
leading order contribution to $\Gamma^a_k ({\bf x})$, which we refer to as bare vertex
$\Gamma^{0, a}_{k} ({\bf x})$, is given by
\begin{widetext}
\be
\label{5-16}
 \Gamma^{0, a}_{k} \lk {{\bf x}}_1, {{\bf x}}_2 ; {\bf y} \rk & = & 
 - \frac{\delta G^{- 1} \lk {{\bf x}}_1, {{\bf x}}_2
 \rk}{\delta A^a_k ({\bf y})} 
 = \frac{\delta}{\delta A^a_k ({\bf y})} \langle 
 {{\bf x}}_1 | g
 \hat{A}^\perp_i \partial_i | {{\bf x}}_2 \rangle
  =  g \frac{\delta}{\delta A^a_k ({\bf y})} \hat{A}^\perp_i ({{\bf x}}_1)
 \partial^{x_1}_i \delta \lk {{\bf x}}_1 - {{\bf x}}_2 \rk \nonumber\\
 & = & g t_{k l} ({\bf y}) \delta ({\bf y} - {{\bf x}}_1) 
 \hat{T}^a \partial^{x_1}_l \delta \lk
 {{\bf x}}_1 - {{\bf x}}_2 \rk  \hk 
 \ee
 or in momentum space 
\begin{align}
\label{5-17}
\Gamma^{0, a}_{k} ({{\bf x}}_1, {{\bf x}}_2; {\bf y}) &= \int \frac{d^3
k}{(2 \pi)^3} \int \frac{d^3 q}{(2 \pi)^3} \Gamma^{0, a}_k ({\bf q}, {\bf k})
e^{i {\bf k} ({\bf y} - {{\bf x}}_1)} \cdot e^{i {\bf q} ({{\bf x}}_1 - {{\bf x}}_2)}
\intertext{by}
\label{5-17a}
\Gamma^{0, a}_k ({\bf q}, {\bf k}) &= i g \hat{T}^a t_{k l} ({\bf k}) q_l \hk .
\end{align}
\bi

\no
\subsection{The ghost self-energy \label{sec3b2}}
\bi

\no
The ghost self-energy $\Sigma$ is conveniently evaluated from its defining equation (\ref{57}) by using
Wick's theorem in the form of eq. (\ref{5-11})
\be
\label{1000}
\Sigma G_\omega = \left\langle g \hat{A}_k \partial_k G \right\rangle_\omega
&=& \left\{ \exp \left[ \frac{1}{4} {\int d^3 x \int d^3 x' \frac{\delta}{\delta
A^{\perp a}_i ({\bf x})} t_{i j} ({\bf x}) \omega^{- 1} \lk {\bf x}, {\bf x}' \rk
\frac{\delta}{\delta A^{\perp a}_j ({\bf x}')}} \right] g \hat{A}_k \partial_k G
\right\}_{A = 0} \hk \nonumber \\ 
&=& \left\{ \left[ \exp \left( \frac{1}{4} {\int d^3 x \int d^3 x' \frac{\delta}{\delta
A^{\perp a}_i ({\bf x})} t_{i j} ({\bf x}) \omega^{- 1} \lk {\bf x}, {\bf x}' \rk
\frac{\delta}{\delta A^{\perp a}_j ({\bf x}')}} \right) , g \hat{A}_k \partial_k \right] G
\right\}_{A = 0} \hk .
\ee
\end{widetext}
Using now the relation
\be
\label{12-5}
\left[ e^{f \lk \frac{\delta}{\delta A} \rk}, A \right] = 
\left[ f \lk \frac{\delta}{\delta A} \rk , A
\right] e^{f \lk
\frac{\delta}{\delta A} \rk}  \hk .
\ee
and expressing $ \frac{\delta}{\delta A} g \hat{A}_k \partial_k $ via eq. (\ref{25})
as $ - \frac{\delta}{\delta A} G^{-1} $ , where the latter quantity represents in view of eq. (\ref{5-16}) 
the bare ghost-gluon vertex $\Gamma^0$ , we find
\begin{widetext}
\be
\int d^3 x_3 \Sigma({{\bf x}}_1,{{\bf x}}_3) G_{\omega}({{\bf x}}_3,{{\bf x}}_2)
&=& \left\{ \frac{1}{2} \int d^3 y_1 \int d^3 y_2 \int d^3 x_3 \hk 
\frac{\delta}{\delta A^{\perp c}_k ({{\bf y}}_1)} 
t_{k l} ({{\bf y}}_1) \omega^{- 1} \lk {{\bf y}}_1, {{\bf y}}_2 \rk \Gamma_l^{0,c} \lk {{\bf x}}_1, {{\bf x}}_3
; {{\bf y}}_2 \rk \right. 
\nonumber\\
& & \left. \exp \left[ \frac{1}{4} {\int d^3 x \int d^3 x' \frac{\delta}{\delta
A^{\perp a}_i ({\bf x})} t_{i j} ({\bf x}) \omega^{- 1} \lk {\bf x}, {\bf x}' \rk
\frac{\delta}{\delta A^{\perp a}_j ({\bf x}')}} \right]  G \lk {{\bf x}}_3,{{\bf x}}_2 \rk
\right\}_{A = 0} \hk .
\ee
Since $\Gamma^0$ and $\omega$ are independent of $A$ the variational derivatives act only
on $G \lk {{\bf x}}_3,{{\bf x}}_2 \rk$ and we obtain
\be
\int d^3 x_3 \Sigma({{\bf x}}_1,{{\bf x}}_3) G_{\omega}({{\bf x}}_3,{{\bf x}}_2)
&=&  \frac{1}{2} \int d^3 y_1 \int d^3 y_2 \int d^3 x_3  \hk
t_{k l} ({{\bf y}}_1) \omega^{- 1} \lk {{\bf y}}_1, {{\bf y}}_2 \rk \Gamma_l^{0,c} \lk {{\bf x}}_1, {{\bf x}}_3
; {{\bf y}}_2 \rk 
\nonumber\\
& & \left\langle \frac{\delta}{\delta A^{\perp c}_k ({{\bf y}}_1)} G \lk {{\bf x}}_3,{{\bf x}}_2 \rk
\right\rangle_{\omega}
\ee
\end{widetext}
where we have again used Wick's theorem (\ref{5-11}). Finally expressing the remaining expectation 
value by means of the defining equation for the ghost-gluon vertex (\ref{y2}) we obtain for the
full ghost self-energy
\begin{widetext}
\be
\label{1002}
\Sigma({{\bf x}}_1,{{\bf x}}_2) = \int d^3 y_1 \int d^3 y_2 \int d^3 x_3 \int d^3 x_4 \hk
D_{kl}^{ab} \lk {{\bf y}}_1,{{\bf y}}_2 \rk \Gamma_l^{0,b} \lk {{\bf x}}_1, {{\bf x}}_3; {{\bf y}}_2 \rk
G_{\omega} \lk {{\bf x}}_3,{{\bf x}}_4 \rk \Gamma_k^{a} \lk {{\bf x}}_4, {{\bf x}}_2; {{\bf y}}_1 \rk
\ee
\end{widetext}
where we have introduced the short hand notation
\be
\label{1003}
D_{kl}^{ab} \lk {{\bf y}}_1,{{\bf y}}_2 \rk =\frac {1}{2} t_{k l} ({{\bf y}}_1)
\omega^{- 1} \lk {{\bf y}}_1, {{\bf y}}_2 \rk \delta^{ab}
\ee
for the gluon propagator. Note that $\Gamma_{k}^{a}$ and $G_{\omega}$ are both off-diagonal
matrices in the adjoint representation. Only the gluon propagator (\ref{1003}), (\ref{5-4}) 
is color diagonal due to our spezific ansatz (\ref{46x}), choosing $\omega$ color independent. 
The ghost self-energy $\Sigma$ (\ref{1002}) is diagrammatically illustrated in figure \ref{10f}.
Investigations (in Landau gauge \cite{R28}) show that vertex dressing is a subleading effect
[36]. Therefore we 
will use in the present paper the so-called rainbow-ladder approximation, replacing the full 
ghost-gluon vertex $\Gamma$ by its bare one $\Gamma^0$ (\ref{5-16}). 
Then the ghost self-energy becomes
\be
\label{1004}
\Sigma^{ab} \lk {{\bf x}}_1,{{\bf x}}_2 \rk &=& \frac{1}{2} \delta^{ab} N_C g^2 t_{k l} ({{\bf x}}_2)
\omega^{- 1} \lk {{\bf x}}_2, {{\bf x}}_1 \rk \nonumber\\
& & \partial_{k}^{x_1} \partial_{l}^{x_2} G_{\omega} \lk {{\bf x}}_1,{{\bf x}}_2 \rk
\ee
where we have used
\be
\label{1005}
\lk \hat{T}^c \hat{T}^c \rk^{a b} = - \delta^{ab} N_C \hk .
\ee

\no
\subsection{The ghost and Coulomb form factor \label{sec3b3}}
\bi

Iterating the Dyson equation (\ref{58}) for the ghost propagator
we end up with a geometric series in powers of $\Sigma
G_0$
\be
\label{59}
G_\omega = G_0 \sli^\infty_{n = 0} \lk \Sigma G_0 \rk^n = 
G_0 \frac{1}{1 - \Sigma
G_0} : = G_0 \frac{d}{g} \hk .
\ee
In the last relation we have introduced the ghost form factor
\be
\label{60}
d = \frac{g}{1 - \Sigma G_0 } \hk .
\ee
\begin{figure}
\includegraphics [scale=0.28] {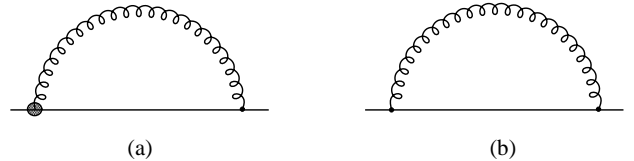}
\caption{\label{10f} Diagrammatic representation of the
ghost self-energy eq. (\ref{1002}). (a) full, (b) in rainbow-ladder approximation.}
\end{figure}
Using the rainbow ladder approximation (\ref{1004}) where $\Sigma$ is given
by the diagramm shown in figure \ref{10f}(b)
the Dyson equation (\ref{58}) 
for the form factor of the ghost
propagator becomes
\be
\label{62}
g d^{- 1} = 1 - \Sigma G_0  : = 1 - g I_d \hk 
\ee
or after Fourier transformation
\be
\label{63}
\frac{1}{d ({\bf k})} &=& \frac{1}{g} - I_d ({\bf k}) \hk , \\
 \hk I_d ({\bf k}) &=& \frac{N_C}{2} \int \frac{d^3 q}{(2 \pi)^3}
\lk 1 - (\hat{{\bf k}} \hat{{\bf q}})^2 \rk \frac{ d ({\bf k}  -
{\bf q})}{({\bf k} - {\bf q} )^2 \omega ({\bf q})} \hk , \nonumber
\ee
%
where $\hat{{\bf k}} = {\bf{k}}/k $. 
Once, the ghost propagator $G$ is known, the Coulomb propagator $F$ can be
obtained from the ghost propagator $G$ by using the relation
\be
\label{12-54}
g \frac{\partial G}{\partial g} = - G + F \hk ,
\ee
which follows immediately from the definitions (\ref{25}), (\ref{13}). This can
be rewritten in a more compact form by 
\be
\label{44}
F =  \frac{\partial}{\partial g} \left( g G \right)  \hk .
\ee
Given the structure of the
Coulomb propagator it is convenient to introduce yet another form factor, which
measures the deviation of the Coulomb propagator
\be
\label{64}
F_\omega := \langle F \rangle_\omega = \langle G (- \partial^2) G \rangle_\omega
\ee
from the factorized form
$\langle G \rangle_\omega (- \partial^2 ) \langle G \rangle_\omega$.
In momentum space we define this form factor $f ({\bf k})$ by 
\be
\label{65}
F_\omega ({\bf k}) & = & G_\omega ({\bf k}) {\bf k}^2 f 
({\bf k}) G_\omega ({\bf k})  \nonumber\\
& = & \frac{1}{g^2} \frac{1}{{\bf k}^2} d ({\bf k}) f ({\bf k}) d ({\bf k}) 
\hk ,
\ee
where we have used eq. (\ref{59}). By taking the expectation value of
(\ref{44}) we have
\be
\label{12-xxx}
F_\omega = \langle \omega | \frac{\partial}{\partial g} (g G) | \omega \rangle \hk .
\ee
Later on $\omega$ will be determined by minimizing the energy. Then $\omega$
becomes $g$-dependent. 
Ignoring this implicit $g$-dependence of $\omega$ we may write 
$\langle \omega | \frac{\partial G}{\partial g} | \omega \rangle = \frac{\partial}{\partial g}
G_\omega$. 
Then from eq. (\ref{12-xxx}), (\ref{65}) and (\ref{59})
follows
\be
\label{66}
f ({\bf k}) =  g^2 d^{- 1} ({\bf k}) \frac{\partial d ({\bf k})}{\partial g} 
d^{- 1} ({\bf k}) 
= - g^2 \frac{\partial}{\partial g} d^{- 1} ({\bf k}) \hk .
\ee
Let us emphasize, that this relation, first obtained in \cite{R27}, is only valid, when the implicit
$g$-dependence of $\omega$ is ignored. Note also, that in the above
equations all Greens functions, form factors and vertices like $G_\omega,
\Sigma, d$ and $f$ are color matrices in the adjoint representation of the gauge
group. \\
In the rainbow ladder approximation defined by eq. (\ref{1004}) the ghost
self-energy $\Sigma$ is given by the one-loop diagramm shown in figure \ref{10f}(b). In
this approximation the ghost form factor (\ref{63}) is a unit matrix in color
space and from eq. (\ref{66}) we find for the Coulomb form factor
\be
\label{13-XXX}
f ({\bf k}) &=& 1 + I_f ({\bf k}), \\
I_f ({\bf k}) &=& \frac{N_C}{2} \int \frac{d^3 q}{(2 \pi)^3} \lk 1 - 
(\hat{{\bf k}}
\hat{{\bf q}})^2 \rk \frac{d ({\bf k} - {\bf q})^2 f 
({\bf k} - {\bf q})}{({\bf k} -
{\bf q})^2 \omega ({\bf q})} \hk . \nonumber
\ee
Here, we have discarded terms involving $\frac{\partial \omega}{\partial g}$ for
reasons explained above. The integral equation (\ref{13-XXX}) is graphically
illustrated in figure \ref{2a}(a). Iteration of this equation yields the
diagrammatic serie shown in figure \ref{2a}(b).
From this equation it is also seen, that in leading order the Coulomb form factor is
given by $f ({\bf k}) = 1$. 
\bi
%
\begin{figure}
\includegraphics [scale=0.32] {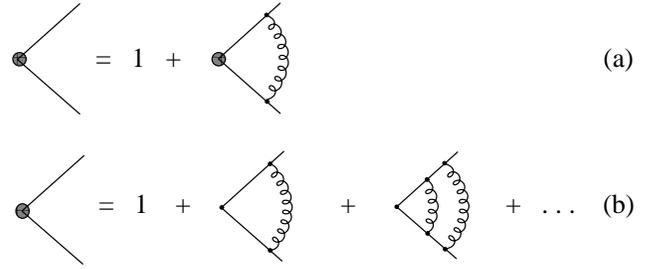}
\caption{ \label{2a} (a) Diagrammatic representation of the integral equation (\ref{13-XXX})
for the Coulomb form factor, (b) Series of diagrams summed up by the integral
equation shown in (a).}
\end{figure}
\bi

\no
\subsection{The curvature \label{sec3c}}
\bi

\no
Consider now the quantity
\be
\label{12-1}
\frac{\delta \ln {\cal{J}}}{\delta A^a_k ({{\bf x}})} & = & \frac{\delta }{\delta A^a_k 
({{\bf x}})} Tr \ln G^{- 1} 
 =  T r \lk G \frac{\delta G^{- 1}}{\delta A^a_k ({{\bf x}})} \rk  . \hs
\ee
Using the definition of the bare ghost-gluon vertex (\ref{5-16}), we have
\be
\label{12-2}
\frac{\delta \ln {\cal{J}}}{\delta A^a_k ({{\bf x}})} = - Tr \lk G \Gamma^{0, a}_{k} 
({{\bf x}}) \rk \hk .
\ee
With this relation the momentum operator (\ref{24}) becomes
\be
\tilde{\Pi}^a_k ({{\bf x}}) = \Pi^a_k ({{\bf x}}) + 
\frac{1}{2 i} Tr \lk G \Gamma^{0, a}_k ({{\bf x}})
\rk \hk .
\ee
Since $\Gamma^0$ (\ref{5-17a}) is independent of the gauge field, we find from
(\ref{12-2})
\be
\label{12-3}
\left\langle \frac{\delta \ln {\cal{J}}}{\delta A^a_k ({{\bf x}})} \right\rangle_\omega = - Tr \lk G_\omega 
 \Gamma^{0, a}_{k} ({{\bf x}}) \rk \hk .
\ee
\begin{figure}
\includegraphics [scale=0.3] {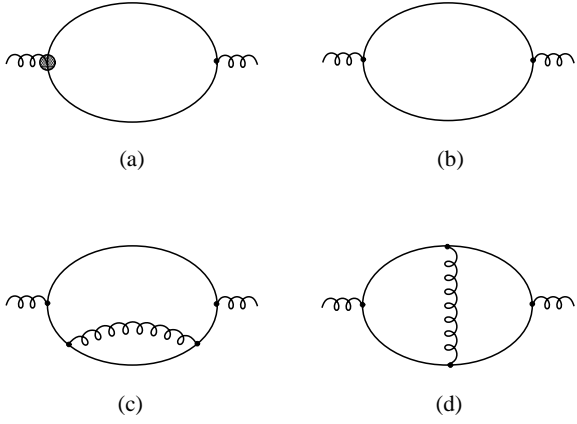}
\caption{\label{68} Diagrammatic representation of the curvature. The line stands for
the full ghost propagator. (a) full curvature given in eq. (\ref{y3}), (b)
results from (a) by replacing the full ghost-gluon vertex $\Gamma$ by the bare
one $\Gamma^{(0)}$, see eq. 
(\ref{12-4}). (c) ghost self-energy correction included in (b). (d) vertex
correction not included in (b). Note, that all vertex corrections contain at
least two loops.}
\end{figure}
Taking a second functional derivative of eq. (\ref{12-2}) and subsequently the
expectation value, and using eq. (\ref{y2}), we obtain
\be
\label{y3}
\left\langle \frac{\delta^2 \ln {\cal{J}}}{\delta A^a_k ({{\bf x}}) \delta A^b_l ({\bf y})} 
\right\rangle_\omega 
& = &  - \left\langle Tr \left[ \frac{\delta G}{\delta A^{\perp a}_k ({\bf x})} \Gamma^{0, b}_l ({\bf y}) \right]
\right\rangle_\omega \nonumber\\
& = & - Tr \left[  \left\langle \frac{\delta G}{\delta A^{\perp a}_k ({\bf x})} 
\right\rangle_\omega \Gamma^{0, b}_l ({\bf y}) 
\right] \nonumber\\
& = &- Tr \left[ G_\omega
\Gamma^a_k ({{\bf x}}) G_\omega \Gamma^{0, b}_l ({\bf y}) \right] \nonumber\\
& = : & - 2 \chi^{a b}_{k l} ({\bf x},{\bf y}) \hk .
\ee
This quantity represents that part of 
the self-energy of the gluon, which is generated by the ghost
loop, and, in the spririt of many-body theory, is referred to as gluon
polarization. It can be also interpreted as that part of the color dielectric
suszeptibility of the Yang-Mills vacuum which originates from the presence of 
the ghost, i.e.
from the curvature of the orbit space. For this reason we will refer to this quantity also as the curvature 
tensor and
the quantity
\be
\label{69x}
\chi ({{\bf x}, {\bf y}}) =  \frac{1}{2} \frac{1}{N^2_C - 1} 
\delta^{a b} t_{k l} ({{\bf x}}) \chi^{ab}_{k l} ({\bf x},{\bf y})
\ee
is referred to as the scalar curvature \footnote{Strictly speaking the true
dimensionless curvature is given  (in momentum space) by $\chi ({\bf k}) / \omega
({\bf k})$. As we will see later, 
it is this quantity, which asymptotically $(k \to
\infty)$ vanishes. We will, however, continue to refer to $\chi$ as curvature.}.
Let us emphasize, that the curvature $\chi ({\bf x},{\bf x}')$ is entirely 
determined
by the ghost propagator $G ({\bf x},{\bf x}')$. Note also, that this quantity contains both
the full (dressed) and the bare ghost-gluon vertices, as illustrated in figure \ref{68}(a). \\
%
As already mentioned above 
from the studies of the Schwinger-Dyson equations in Landau gauge \cite{R28} 
it is known, that the effect of the vertex dressing is subleading \footnote{ \label{f1} To
be more precise in this gauge the ghost-gluon vertex is not renormalized, so
that the bare vertex has the correct ultraviolet behaviour. Furthermore, the
effect of the dressing of the vertex is small in the infrared \cite{R28}.}. 
We will
therefore replace the full vertex $\Gamma$ by the bare one (\ref{5-16}) (rainbow-ladder approximation). 
The curvature is then given by the diagram shown in figure \ref{68}(b).
Using the
explicit form of the bare ghost-gluon vertex (\ref{5-16}), 
the curvature tensor (ghost loop part of the gluon polarization) becomes
%
\begin{widetext}
\vspace{-0.2cm}
\be
\label{12-4}
\chi^{a b}_{k l} ({\bf x},{\bf y}) =  \frac{1}{2} g^2  t_{k m} ({{{\bf x}}}) 
t_{l n} ({{\bf y}}) Tr \left[ \hat{T}^a \lk \partial^x_m G_\omega 
({\bf x},{\bf y}) \rk \hat{T}^b
\lk \partial^y_n G_\omega ({\bf y},{\bf x}) \rk \right] \hk .
\ee
\end{widetext}
In the rainbow ladder approximation used in this paper the ghost Greens function
is a unit matrix in color space. Using $tr ( \hat{T}^a \hat{T}^b ) = - \delta^{a b}
N_C$ and $\chi ({\bf x},{\bf y}) = \chi ({\bf x }-{\bf y})$ 
we obtain
\vspace{-0.3cm}
\be
\label{71b}
t_{k n} \lk {\bf x} \rk \chi_{nl}^{a b} \lk {\bf x} , {\bf y} \rk  
& = & \delta^{a b} t_{k l} \lk {\bf x} \rk
\chi \lk {\bf x} , {\bf y} \rk
\ee
with the scalar curvature (\ref{69x}) in momentum space given by 
\be
\label{71}
\chi ({\bf k})  & = & \frac{N_C}{4} \int \frac{d^3 q}{(2 \pi)^3}  
\lk 1 - (\hat{{\bf k}}
\hat{{\bf q}})^2 \rk \frac{d ({\bf k} -{\bf q}) d ({\bf q})}{({\bf k} - {\bf q})^2} \hk . \hs
\ee
For the evaluation of the vacuum energy to be carried out in the next section we
also need the quantity $\langle A^\perp \frac{\delta \ln {\cal{J}}}{\delta A} \rangle$. This
expectation value can be conveniently calculated using the Wick's theorem in the
form of eq. (\ref{5-11}) and the relation (\ref{12-5}).
We then obtain
\be
\left\langle A^{\perp a}_k ({\bf x}') \frac{\delta \ln {\cal{J}}}{\delta A^b_l ({\bf x})} 
\right\rangle_\omega & = & \frac{1}{2} \int d^3 x_1
t_{k m} ({\bf x}') \omega^{- 1} \lk {\bf x}', {\bf x}_1 \rk \hs \hk \nonumber\\
& & \cdot \left\langle \frac{\delta}{\delta A^a_m ({\bf x}_1)} \frac{\delta}{\delta A^b_l ({\bf x})} \ln 
{\cal{J}} \right\rangle_\omega .
\ee
With the definition of the curvature $\chi$ (\ref{y3}) this quantity can be
written as
\be
\label{73x}
& & \left\langle A^{\perp a}_k ({{\bf x}'}) \frac{\delta \ln {\cal{J}}}{\delta A^b_l ({{\bf x}})} 
\right\rangle_\omega  = 
\nonumber\\
& & \hspace{1.0cm} - \int d^3 x_1 t_{k m} ({{\bf x}'}) \omega^{- 1} \lk {\bf x}',{{\bf x}}_1 \rk \chi^{a b}_{m l} \lk 
{{\bf x}}_1, {{\bf x}} \rk \hk . \hspace{1.0cm}
\ee
We have now all ingredients available to evaluate the vacuum energy.
\bi

\no
\section{The vacuum energy density \label{sec4}}
\bi

\no
For the Gaussian wave functional it is straightforward to evaluate the
expectation value of the potential (magnetic term) of the Hamilton operator.
Using Wick's theorem, one finds
\begin{widetext}
\vspace{-0.5cm}
\be
\label{5-6}
E_p & = & \langle \omega | H_B | \omega \rangle 
 =  \frac{N^2_C - 1}{2} \delta ({\bf 0}) \int d^3 k \frac{{\bf k}^2}{\omega ({\bf k})}
\hk \hk + \hk \hk \frac{N_C \lk N^2_C - 1 \rk }{16} g^2 \delta ({\bf 0}) \int \frac{d^3 k d^3
k'}{(2 \pi)^3} \frac{1}{\omega ({\bf k}) \omega ({\bf k}')}  \lk 3 - (\hat{{\bf k}} 
\hat{{\bf k}'})^2 \rk \hk . \hs
\ee
\end{widetext}
%
In the above expression for the vacuum energy the divergent constant $\delta ({\bf 0})$ 
can be interpreted as the volume of space. Removing 
this constant we obtain the energy density, which is the more relevant quantity
in quantum field theory. \\
%
The evaluation of the expectation value of the kinetic term and of the Coulomb
term is more involved.
Consider the action of the momentum operator $\tilde{\Pi}$ (\ref{24}) on the
wave functional (\ref{46x})
\begin{widetext}
\be
\label{x1}
\tilde{\Pi}^a_i ({{\bf x}}) \tilde{\Psi} [A^\perp] = \frac{1}{i} t_{i k} ({{\bf x}}) \left[
\frac{\delta}{\delta A^a_k ({{\bf x}})} - 
\frac{1}{2} \frac{\delta \ln {\cal{J}}}{\delta A^a_k
({{\bf x}})} \right] \tilde{\Psi} [A^\perp]  
=  {i} Q^a_i ({{\bf x}}) \tilde{\Psi} [A^\perp] \hk ,
\ee
where
\be
\label{x2}
Q^a_i ({{\bf x}})  =  \int d^3 x_1 \omega ({\bf x},{{\bf x}}_1) A^{\perp a}_i 
({{\bf x}}_1) + \frac{1}{2} t_{i k} ({{\bf x}})
\frac{\delta \ln {\cal{J}}}{\delta A^a_k ({{\bf x}})}
 =  \int d^3 x_1 \omega ({\bf x},{{\bf x}}_1) A^{\perp a}_i ({{\bf x}}_1) - \frac{1}{2}
t_{i k} ({\bf x}) Tr \lk G
\Gamma^{0, a}_{k} ({{\bf x}}) \rk  . \hspace{0.2cm}
\ee
\end{widetext}
The kinetic energy (\ref{24}) can then be expressed as
\be
\label{x3}
E_k = \frac{1}{2}  \int d^3 x \left\langle Q^a_i ({{\bf x}}) Q^a_i ({{\bf x}})
\right\rangle_\omega \hk .
\ee
Analogously the Coulomb energy becomes
\begin{widetext}
\be
\label{x4}
E_c = - \frac{g^2}{2} \int d^3 x \int d^3 x' \left\langle Q^c_i ({{\bf x}})
\hat{A}^{\perp {ca}}_i ({{\bf x}}) F^{a b} ({\bf x},{\bf x}') \hat{A}^{\perp {b d}}_j 
({{\bf x}'}) Q^d_j
({{\bf x}'}) \right\rangle_\omega \hk .
\ee
\end{widetext}
Due to the presence of the curvature term (i.e. the second term) 
in (\ref{x2}) the vacuum expectation
value $\langle \dots \rangle_\omega$ cannot be worked out in closed form since ${\cal{J}} [A]$ is an
infinite series in the gauge potential. For practical purposes we have to resort
to approximations. Throughout the paper we shall evaluate propagators to one-loop
level and accordingly the vacuum energy to two-loop level. 
To this order the following factorization holds
in the Coulomb term
\begin{align}
\label{x5}
\langle Q A^\perp F A^\perp Q \rangle_\omega = \langle F \rangle_\omega \left[ \langle A^\perp A^\perp \rangle_\omega \langle
Q Q \rangle_\omega 
+ \langle A Q \rangle_\omega \langle A Q \rangle_\omega \right] \hk ,
\end{align}
where we have used
\be
\label{x5a}
\langle A\rangle_\omega = 0 \hk , \hk \langle Q \rangle_\omega = 0 \hk .
\ee
Furthermore in $\langle Q Q \rangle_\omega$ only a single $A$ from one $Q$ has to be contracted
with an $A$ from the other $Q$. The remaining $A$'s of a $Q$ have to be
contracted among themselves. All other contractions give rise to diagrams
with more than two loops. This ensures, that $\langle Q Q \rangle_\omega$ remains a total
square even after the vacuum expectaion value is taken. This fact can be
utilized to simplify the evaluation of $\langle Q Q \rangle_\omega$ \footnote{Note, that inside the
expectation values $\langle Q Q \rangle_\omega$ and $\langle A Q \rangle_\omega$ the quantity  $Q$ 
has the structure $Q
= \int S A$, where $S ({\bf x},{\bf x}') = \delta ({{\bf x} -{\bf x}'}) - 
\int d x_1 \omega^{- 1} ({\bf x},
{{\bf x}}_1) \chi ({{\bf x}}_1, {{\bf x}'})$, 
so that the above expectation values have to be taken to
one-loop order as
$$
\langle Q Q \rangle  =  \langle S A S A \rangle = \langle S \rangle \langle S 
\rangle \langle A A \rangle  
$$
and
$$
\langle A Q \rangle  =  \langle S \rangle \langle A A \rangle \hk .
$$
}.
To obtain $\langle Q Q \rangle_\omega$ we need
explicitly to evaluate only the first term $\sim \langle A A \rangle_\omega $ and the mixed term
$\sim 2 \langle A \frac{\delta \ln {\cal{J}}}{\delta A}\rangle_\omega$. 
The remaining term $\langle \frac{\delta \ln
{\cal{J}}}{\delta A} \frac{\delta \ln
{\cal{J}}}{\delta A} \rangle_\omega$ can be found by quadratic completion. Using eq. (\ref{73x}) 
 we find for the expectation value of the mixed term
\begin{widetext}
\be
\label{x6}
\left\langle A^{\perp a}_i ({{\bf x}}) Q^b_j ({\bf x}') \right\rangle_\omega & = & 
 \frac {1}{2} \delta^{a b} t_{i j} ({{\bf x}})
 \left[ \delta ({{\bf x}
- {\bf x}'}) - \int d^3 x_1 \omega^{- 1} ({\bf x},{{\bf x}}_1) \chi ({{\bf x}}_1, {\bf x}') 
\right] \hk ,
\ee
\end{widetext}
where $\chi ({\bf x}, {\bf x}')$ is the scalar curvature of the gauge orbit space defined by
eq. (\ref{69x}). With this result one then finds by quadratic completion
\begin{widetext}
\be
\label{x7}
\left\langle Q^a_i ({{\bf x}}) Q^b_j ({{\bf x}'}) \right\rangle_\omega  = \frac {1}{2}
 \delta^{ab} \int d^3 x_1 \int d^3
 x_2 \left[ \omega ({\bf x},
{{\bf x}}_1) - \chi ({\bf x}, {{\bf x}}_1) \right] 
 t_{i j} ({{\bf x}}_1) \omega^{- 1} ({{\bf x}}_1, {{\bf x}}_2) \left[ \omega 
({{\bf x}}_2, {{\bf x}'}) - \chi ({{\bf x}}_2, {{\bf x}'})
\right] \hk .
\ee
The kinetic energy (\ref{x3}) becomes then with $\delta^{a a} = N^2_c - 1$ and
$t_{i i} ({{\bf x}}) = 2$
\be
\label{x8}
E_k & = & \frac {\lk N^2_c - 1 \rk}{2} \int d^3 x \int d^3 x_1 \int d^3 x_2 \left[
\omega ({{\bf x}},{{\bf x}}_1) - \chi ({{\bf x}},{{\bf x}}_1) \right] 
 \omega^{- 1} ({{\bf x}}_1, {{\bf x}}_2) \left[ \omega ({{\bf x}}_2, {\bf x}) - 
\chi ({{\bf x}}_2, {\bf x}) \right] \hk .
\ee
Analogously, one finds for the Coulomb energy using (\ref{x5}), (\ref{x5a}), 
(\ref{x6}),(\ref{x7})  and $f^{a b c} f^{a b d } = N_C
\delta^{c d}$
\be
\label{x9}
E_c & = & \frac {N_C \lk N^2_c - 1 \rk} {8} g^2 \int d^3 x \int d^3 x' F_\omega
 ({{\bf x}, {\bf x}'})
\biggl[ \left[ t_{i j} ({{\bf x}}) \omega^{- 1} ({{\bf x}, {\bf x}'}) \right] \nonumber\\
& & \int d^3 x_1 \int d^3 x_2 \left[ \omega ({\bf x}, {{\bf x}}_1) - \chi 
({\bf x}, {{\bf x}}_1) \right] \left[ t_{j i}
({{\bf x}}_1) \omega^{- 1} ({{\bf x}}_1, {{\bf x}}_2) \right] \left[ \omega 
({{\bf x}}_2, {{\bf x}'}) - \chi ({{\bf x}}_2, {{\bf x}'}) \right]
\nonumber\\
& & - \lk t_{i j} ({{\bf x}}) \left[ \delta ({{\bf x}} - {{\bf x}'}) - \int d^3 x_1 
\omega^{- 1} ({{\bf x}}, {{\bf x}}_1)
\chi ({{\bf x}}_1, {{\bf x}'}) \right] \rk 
  \lk t_{j i} ({{\bf x}'}) \left[ \delta ({{\bf x}'} - {{\bf x}}) - \int d^3 
x_2 \omega^{- 1} ({{\bf x}'}, {{\bf x}}_2)
\chi ({{\bf x}}_2, {{\bf x}}) \right] \rk \biggr] . \hs \hk
\ee
These expressions can be written in a more compact form in momentum space:
\be
\hspace{-4cm} E_k = \frac{N^2_C - 1}{2} \delta^{(3)} ({\bf 0}) \int d^3 k \frac{ \left[ \omega ({\bf k}) - \chi
({\bf k}) \right]^2}{\omega ({\bf k})}
\ee
and
\be
E_c & = & \frac{N_C (N^2_C - 1)}{8} \delta^{(3)} ({\bf 0}) \int \frac{d^3 k d^3 k'}{(2
\pi)^3} \lk 1 + (\hat{{\bf k}} \hat{{\bf k}'})^2 \rk \frac{d ({\bf k} -{\bf k}')^2 f
({\bf k} - {\bf k}')}{({\bf k} - {\bf k}')^2} \nonumber\\
& & \cdot \frac{\Big( \left[ \omega ({\bf k}) - \chi ({\bf k}) \right] - 
\left[ \omega ({\bf k}') - \chi ({\bf k}') \right] \Big)^2}{\omega ({\bf k}) \omega ({\bf k}')} \hk ,
\ee
\end{widetext}
where again the divergent factor $\delta ({\bf 0})$ has to be removed to obtain the
corresponding energy densities. \\
%
The kernel $\omega ({\bf k})$ in the Gaussian ansatz of the wave functional is
determined from the variational principle $\delta E (\omega)  / \delta \omega =
0$. Variation of the magnetic energy is straightforward and yields
\begin{widetext}
\be
\label{17-1}
\frac{\delta E_p}{\delta \omega ({\bf k})} = - \frac{N^2_C - 1}{2} \delta^{(3)} ({\bf 0})
\frac{1}{\omega ({\bf k})^2} \left[ {\bf k}^2 + \frac{N_C}{4} g^2 \int \frac{d^3 q}{(2
\pi)^3} \lk 3 - (\hat{{\bf k}} \hat{{\bf q}} )^2 \rk \cdot \frac{1}{\omega ({\bf q})} \right]
\hk .
\ee
\end{widetext}
The kinetic and Coulomb part of the vacuum energy $E_k, E_c$ do not only
explicitly depend on the kernel $\omega ({\bf k})$, but also implicitly via the ghost
form factor $d ({\bf k})$ and the Coulomb form factor $f ({\bf k})$ as well as via the
curvature $\chi ({\bf k})$. However, one can show, that variation of these quantities
$d ({\bf k}), f ({\bf k}), \chi ({\bf k})$ with respect to $\omega ({\bf k})$ gives rise to two-loop
terms, which are beyond this cope of this paper. Ignoring these terms, we obtain
for the variation of the kinetic and Coulomb parts of the energy
\begin{widetext}
\be
\label{17-2}
\frac{\delta E_k}{\delta \omega ({\bf k})} & = & \frac{N^2_C - 1}{2} \delta^{(3)} ({\bf 0}) \left[ 1 -
\frac{\chi ({\bf k})^2}{\omega ({\bf k})^2} \right] \hk ,
\\
\label{17-3}
\frac{\delta E_c}{\delta \omega ({\bf k})} & = & \frac{N_C (N_C^2 - 1)}{8} \delta^{(3)} ({\bf 0})
\frac{1}{\omega ({\bf k})^2} 
 \cdot \int \frac{d^3 q}{(2 \pi)^3} \lk 1 + (\hat{{\bf k}} \hat{{\bf q}} )^2 \rk
\frac{d ({\bf k} - {\bf q})^2 f ({\bf k} - {\bf q})}{({\bf k} - {\bf q})^2} \cdot \frac{\omega ({\bf k})^2
- \left[ \omega ({\bf q}) - \chi ({\bf q}) + \chi ({\bf k}) \right]^2}{\omega ({\bf k})} \hk . \hs \hs
\ee
\end{widetext}
Putting all these ingredients together, we find from the variational principle
the gap equation to one-loop order 
%
\be
\label{17-4}
\omega ({\bf k})^2 = {\bf k}^2 + \chi ({\bf k})^2   
+ I_\omega ({\bf k}) + I^0_\omega \hk ,
\ee
where we have introduced the abbreviations
\begin{widetext}
\be
\label{17-5a}
I^0_\omega & = & \frac{N_C}{4} g^2 \int 
\frac{d^3 q}{(2 \pi)^3}
\lk 3 - (\hat{{\bf k}} \hat{{\bf q}})^2 \rk \frac{1}{\omega ({\bf
q})} \hk , \\
\nonumber\\
\label{17-5}
I_\omega ({\bf k}) & = & \frac{N_C}{4} \int \frac{d^3 q}{(2 \pi)^3} \lk 1 + (\hat{{\bf k}} \hat{{\bf q}})^2
\rk 
 \cdot \frac{d ({\bf k} -{\bf q})^2 f ({\bf k} - {\bf q})}{({\bf k} - {\bf q})^2} 
\cdot \frac{\left[ \omega ({\bf q}) - \chi ({\bf q}) + \chi
({\bf k}) \right]^2 - \omega ({\bf k})^2}{\omega ({\bf q})} \hk .
\ee
\end{widetext}
%
We are then left
with a set of four coupled Schwinger-Dyson equations for the ghost
form factor $d (k)$ (\ref{63}),
for the curvature $\chi (k)$ (\ref{71}),
for the frequency $\omega (k)$ (gap equation) (\ref{17-4})
and for the Coulomb form factor $f (k)$ (\ref{13-XXX}). Before
 finding the self-consistent solutions
to these coupled Schwinger-Dyson equations in the next section we shall 
study their analytic properties
 in the ultraviolet $(k \to \infty)$
and in the infrared $(k \to 0)$.
\bi

\no
\section{Asymptotic behaviour \label{sec5}}
\bi

\no
To obtain first insights into the
infrared and ultraviolet 
behaviour of solutions of the coupled Schwinger-Dyson equation, 
we investigate these
 equations in the so-called angular approximation, which is defined in eq.
 (\ref{X1}) below. At this stage we have no estimate of the accuracy of this
 approximation. However, the numerical calculation to be presented in section \ref{sec7},
 which will be carried out without resorting to the angular approximation, will
 produce asymptotic behaviours, which are quite similar to the one obtained
 below within the angular approximation. \\
The angular approximation is defined
by approximating a function $h ({\bf k} - {\bf q})$, 
which occurs under the momentum integral
with an argument given by the difference between the external momentum ${\bf k}$
and the internal loop 
momentum ${\bf q}$ integrated over, by the following expression
\be
\label{X1}
h (| {\bf k} - {\bf q} |) = h (k) \Theta (k - q) + h (q) \Theta (q - k) \hk .
\ee
With this approximations the angle integrals in the Schwinger-Dyson equations
for $\chi (k)$ and $d (k)$
reduce with $\lk 1 - (\hat{k} \hat{q})^2 \rk = \sin^2 \vartheta$ 
to the following integrals
\begin{widetext}
\be
\label{X2}
\il^\pi_0 d \vartheta \sin^3 \vartheta \frac{d ({\bf k} - {\bf q})}{({\bf k} - {\bf q})^2} 
& \simeq &
\left[ \Theta (k - q) \frac{d (k)}{k^2}  + 
\Theta (q - k) \frac{d (q)}{q^2} \right] \cdot \il^\pi_0 d \vartheta \sin^3 \vartheta  \nonumber\\
& = & \frac{4}{3} \left[ \Theta (k - q) \frac{d (k)}{k^2} + \Theta (q - k)
\frac{d (q)}{q^2} \right] \hk .
\ee
\end{widetext}
Using these results the remaining integrals over the modulus of the momentum
occuring in the Schwinger-Dyson equations become
\be
\label{XZ1}
I_d (k) & = & \frac{N_C}{6 \pi^2} \left[ \frac{d (k)}{k^2} \il^k_0 d q
\frac{q^2}{\omega (q)} + \il^\Lambda_k d q \frac{d (q)}{\omega (q)} \right] ,
\\
\label{XZ2}
I_\chi (k) & = & \frac{N_C}{12 \pi^2} \left[ \frac{d (k)}{k^2} \il^k_0 d q q^2 d
(q) + \il^\Lambda_k d q d (q)^2 \right]   \hk . \hk \hk \hk \hk \hk \hk
\ee
It turns out, that the simplest way to solve the Schwinger-Dyson
 equation for the ghost
propagator and the curvature in the asymptotic regions 
is to differentiate these equations with respect to
$k$. The clue is, that the derivatives contain only ultraviolet convergent
integrals.  Indeed from equations (\ref{XZ1}), (\ref{XZ2}) we have 
(The contributions from the derivative of the integration
limits $k$ cancel)
\be
\label{XX1}
I'_d (k) & = &  \frac{N_C}{6 \pi^2} \frac{1}{k^2} \left[ d' (k) - 2 \frac{d
(k)}{k} \right] \il^k_0 d q \frac{q^2}{\omega (q)} \\
\label{XX2}
I'_\chi (k) & = &  \frac{N_C}{12 \pi^2} \frac{1}{k^2} \left[ d' (k) - 2 \frac{d
(k)}{k} \right] \il^k_0 d q {q^2} d (q)  \hk .  \hk  \hk
\ee
Differentiating the Schwinger-Dyson equation for the ghost form factor, eq.
(\ref{63}), with
respect to the external momentum $k$ and using equation (\ref{XX1}), we find
\be
\label{19-2}
d' (k) \left[ \frac{1}{d (k)^2} - \frac{N_C}{6 \pi^2} \frac{R (k)}{k^2}
\right] = - \frac{N_C}{3 \pi^2} \frac{R (k)}{k^2} \frac{d (k)}{k} \hk ,
\ee
where we have introduced the abbreviation 
\be
\label{19-3}
R (k) = \il^k_0 d q  \frac{q^2}{\omega (q)} \hk .
\ee
Differentiating  the equation for the
curvature (\ref{71}) and using the angular approximation (\ref{XZ2}) we obtain
\be
\label{111}
\chi' (k) = \frac{N_C}{12 \pi^2} \frac{1}{k^2} \left[ d' (k) - 2 \frac{d (k)}{k}
\right] S (k) \hk ,
\ee
where
\be
\label{GXX}
S (k) = \il^k_0 d q q^2 d (q) \hk .
\ee
Below we will solve these equations separately for $k \to \infty$ and $k \to
0$, respectively.
\bi

\no
\subsection{Ultraviolet behaviour \label{sec5a}}
\bi

\no
Due to asymptotic freedom the gluon energy $\omega (k)$ has to approach for $k
\to \infty$ the asymptotic form
\be
\label{19-1}
\omega (k) \to \sqrt{{\bf k}^2} \hk , \hk k \to \infty \hk .
\ee
This behaviour will also be obtained later on from the solution to the gap
equation. Let us therefore concentrate on the asymptotic behaviour of the ghost
form factor $d (k)$, the curvature $\chi (k)$ and the Coulomb form factor $f
(k)$ resorting to the angular approximaton eq. (\ref{X2}). \\
%
Consider first the Schwinger-Dyson-equation (\ref{19-2}) for the ghost form
factor. It contains the so far unknown integral $R (k)$ (\ref{19-3}).
%
For large $k$ the dominating contribution to this integral 
comes from the large $q$ region.
Hence, for an estimate of $R (k)$ we can use for $\omega (k)$ its
asymptotic value $\omega (q) = \sqrt{{\bf q}^2}$ yielding
\be
\label{19-4}
R (k) = \frac{k^2}{2} \hk , \hk k \to \infty \hk .
\ee
Inserting this result into eq. (\ref{19-2}) we obtain
\be
\label{19-5}
d' (k) \left[ \frac{1}{d (k)^3} - \frac{N_C}{12 \pi^2} \frac{1}{d (k)} \right]
= - \frac{N_C}{6 \pi^2} \frac{1}{k} \hk .
\ee
Integrating this equation yields
\be
\label{106}
\lk \frac{1}{d (k)} \rk^2 = \lk \frac{1}{d (\mu)} \rk^2 + \frac{N_C}{6 \pi^2}
\lk \ln \frac{k^2}{\mu^2} - \ln \frac{d (k)}{d (\mu)} \rk \hk ,
\ee
where $\mu$ is an arbitrary momentum.
%
For large $k$ we expect, that the following condition holds
\be
\label{19-7}
\frac{d (k)}{d (\mu)} << \frac{k^2}{\mu^2} \hk , \hk k \to \infty \hk .
\ee
The reason is, that for $k \to \infty$ the Faddeev-Popov determinant approaches
that of the Laplacian and accordingly the ghost form factor should approach one,
$d (k \to \infty) \to 1$. However, due to dimensional transmutation giving rise to
anomalous dimensions, we will shortly see, that $d (k)$ behaves asymptotically
like $d (k) \sim (\ln k^2 /\mu^2)^{- \frac{1}{2}}$, which of course satisfies the
condition eq. (\ref{19-7}). In anticipation of this result, we will in the
following assume, that the condition eq. (\ref{19-7}) is fullfilled and
subsequently show, that the resulting solution for $d (k)$ will indeed obey this
condition. When this condition is fullfilled, we can ignore the last term in eq.
(\ref{106}) resulting in the explicit solution
\be
\label{19-8}
d (k) = \frac{d (\mu)}{\sqrt{1 + \frac{N_C}{6 \pi^2} d (\mu)^2 \ln \lk
\frac{k^2}{\mu^2} \rk }} \hk ,
\ee
which asymptotically behaves like
\be
\label{19-9}
d (k) = \pi \sqrt{\frac{6}{N_C}} \lk \ln \frac{k^2}{\mu^2} \rk^{- \frac{1}{2}}
\hk 
\ee
as anticipated before.  \\
%
In passing we note, that the obtained asymptotic behaviour of $d (k)$ is
precisely that of the running coupling constant $g \sim \sqrt{\alpha}$. 
In section \ref{sec6} we will see,
that indeed asymptotically $d (k \to \infty)$ 
approaches the renormalized coupling
constant. From eq. (\ref{59}) then follows that the ghost behaves, indeed, asymptotically
like a free particle. \\

From the asymptotic form of the ghost form factor (\ref{19-8}) we can
immediately infer the asymptotic form of the Coulomb form factor $f (k)$ by
using the relation (\ref{66}). This yields
\be
\label{19-10}
f (k) = f (\mu) \frac{d (k)}{d (\mu)} \hk , \hk k \to \infty \hk .
\ee
Thus, the asymptotic behaviour of the Coulomb form factor is up to a numerical
constant $f (\mu) / d (\mu)$ the same as the one of the ghost form factor. \\

The extraction of the asymptotic form of the curvature is somewhat more
involved. 
%
For large $k \to \infty$ we can use the asymptotic form (\ref{19-9}) 
of the ghost form factor
$d (k)$ from which one finds
\be
\label{118}
d' (k) = - \frac{N_C}{6 \pi^2} \frac{1}{k} d (k)^{3} \hk ,
\ee
so that eq. (\ref{111}) becomes
\be
\label{GZZ1}
\chi' (k) = - \frac{N_C}{6 \pi^2} \frac{d (k)}{k^3} \left[ 1 + \frac{N_C}{12 \pi^2}
d (k)^2 \right] S (k) \hk .
\ee
The ghost form factor $d (k)$ is, by definition, strictly positive definite
inside the first Gribov horizon, to which we have to restrict our gauge orbits.
Thus the integral 
$S (k)$ (\ref{GXX}) is positive definite and $\chi' (k) \leq 0$.  For
sufficiently large $k$, where we can use the asymptotic form (\ref{19-9}) 
of $d (k)$ for which 
the integrand in $S (k)$ (\ref{GXX}) 
behaves like $q^2 / \sqrt{\ln q/\mu}$. Furthermore,
we will show in the next section that $q^2 d (q) \to 0$ for $q \to
0$ \footnote{In fact we will later see, that $q^2 d (q)$ is a monotonically
raising function.}.
Therefore for sufficiently large $k$ an upper limit to $S (k)$ is given by
\be
S (k) \leq \il^k_0 d q q^2 = \frac{1}{3} k^3 \hk .
\ee
From eq. (\ref{GZZ1}) we obtain then the following estimate
\be
0 > \chi' (k) \gtrsim - \frac{N_C}{18 \pi^2} d (k) \left[ 1 + \frac{N_C}{12 \pi^2} d
(k)^2 \right]  \hk .
\ee
Since $d (k) \sim 1 /\sqrt{\ln k/\mu}$ for $k \to \infty$ the second term in the
bracket is irrelevant for sufficiently large $k$. By the same token we can
multiply this term by a factor $(- 2)$ without changing the asymptotic result.
Using (\ref{118})
\be
d (k) \left[ 1 - 2 \frac{N_C}{12 \pi^2} d (k)^2 \right] = \frac{d}{d k} \left[ k d (k) \right]
\ee
we obtain the asymptotic estimate
\be
0 > \chi' (k)  \gtrsim - \frac{N_C}{18 \pi^2} \frac{d}{d k} \left[ k d (k) \right] \hk .
\ee
Integrating this equation over the interval $(k_0, k)$ we find
\be
0 > \chi (k) - \chi (k_0)  \gtrsim - \frac{N_C}{9 \pi^2} \left[ k d (k) - k_0 d (k_0) \right] \hk .
\ee
Dividing this equation by $k$ we find that asymptotically
\be
0 > \frac{\chi (k)}{k} \gtrsim - \frac{N_C}{18 \pi^2} d (k) \hk , \hk k \to \infty
\hk 
\ee
or with eq. (\ref{19-9})
\be
\chi (k) \sim \frac{k}{\sqrt{\ln k/\mu}} \hk , \hk k \to \infty  \hk .
\ee
From this relation it follows, that
\be
\frac{\chi (k)}{\omega (k)} \sim  \frac{\chi (k)}{k} \sim \frac{1}{\sqrt{\ln k /\mu}}
\to 0 \hk , \hk  k \to \infty \hk
.
\ee
To summarize, we have found 
the following ultraviolet behaviour $(k \to \infty)$
\be
\label{G17}
\omega ({\bf k}) & \sim & \sqrt{{\bf k}^2} \nonumber\\
d ({\bf k}) & \sim &  \frac{1}{\sqrt{\ln k /\mu}} \hk , \hk f ({\bf k}) \sim
 \frac{1}{\sqrt{\ln k /\mu}}
 \nonumber\\
\frac{\chi ({\bf k})}{\omega ({\bf k})} & \sim &  \frac{1}{\sqrt{\ln k /\mu}} \hk .
\ee
The first equation means, that gluons behave asymptotically $(k
\to \infty)$
like free particles with energy $k$, while the last equation implies,
that the space of gauge connections becomes asymptotically flat. The ghost form
factor $d (k)$ deviates from that of a free (massless) point like particle by
the anomalous dimensions $1/\sqrt{\ln k /\mu}$. These relations
are in accord with asymptotic freedom. \\

One easily convinces oneself that the asymptotic behaviour obtained above 
yields indeed a
consistent solution to the coupled  Schwinger-Dyson equation.
\bi

\no
\subsection{The infrared behaviour \label{sec5b}}
\bi

\no
In the following we study the behaviour of the solutions of the coupled
Schwinger-Dyson equations for $k \to 0$ thereby using again the angular
approximation (\ref{X1}). \\
%
For the quantities under interest in the infrared we make the following
ans\"atze
\be
\label{X5}
\omega (k) = \frac{A}{k^\alpha} \hk , \hk d (k) = \frac{B}{k^\beta} \hk, \hk
\chi (k) = \frac{C}{k^\gamma} \hk .
\ee
With these ans\"atze we solve (the derivatives of) 
the coupled Schwinger-Dyson equations (\ref{19-2}), (\ref{111}) for $k \to 0$. \\
%
In the remaining integrals (\ref{19-3}), (\ref{GXX}), 
the integration variable is restricted to the intervall $0 <
q < k$. For $k \to 0$ we can use the asymptotic representations, eq.
(\ref{X5}), in the integrands and obtain
\be
R (k) & = & \frac{1}{A} \il^k_0 d q q^{\alpha + 2} = \frac{1}{A} \frac{k^{\alpha
+ 3}}{\alpha + 3} \hk , \nonumber\\
S (k) & = & B \il^k_0 d q q^{2 - \beta} = B \frac{k^{3 - \beta}}{3 - \beta} \hk
.
\ee
Inserting these expressions into eqs. (\ref{XX1}), (\ref{XX2}) we find
\be
\label{X7}
I'_d (k) & = & - \frac{N_C}{6 \pi^2} \frac{B}{A} \cdot \frac{\beta + 2}{\alpha +
3} k^{\alpha - \beta } \hk , \nonumber\\
I'_\chi (k) & = & - \frac{N_C}{12 \pi^2} B^2 \frac{\beta + 2}{3 - \beta} k^{- 2
\beta}
 \hk .
\ee
From the derivative of the ghost equation
\be
\label{X8}
\frac{d' (k)}{d (k)^2} = I'_d (k)
\ee
we then obtain the following relation
\be
\label{X9}
\frac{A}{B^2} = \frac{N_C}{6 \pi^2}  \frac{\beta + 2}{\beta (\alpha + 3)} 
k^{\alpha - 2 \beta + 1}
\hk .
\ee
The left-hand side of this equation is constant. The same has to be true for the
right-hand side, which implies
\be
\label{X10}
\alpha = 2 \beta - 1 \hk .
\ee
Inserting this relation into eq. (\ref{X9}) we obtain a relation between the
coefficients $A$ and $B$
\be
\label{X11}
\frac{A}{B^2} = \frac{N_C}{6 \pi^2} \frac{\beta + 2}{2 \beta (\beta + 1)} \hk .
\ee
\bi

\no
Analogously we find with (\ref{X7}) 
from the derivative of the curvature equation
\be
\label{X12}
\chi' (k) = I'_\alpha (k)
\ee
the relation
\be
\label{X13}
\frac{C}{B^2} = \frac{N_C}{12 \pi^2} \frac{\beta + 2}{\gamma (3 - \beta)} 
k^{\gamma
- 2 \beta + 1} \hk .
\ee
Since the left hand side is a constant (i.e. independent of $k$) it follows
\be
\label{X14}
\gamma = 2 \beta - 1 \hk 
\ee
and thus
\be
\label{X15}
\frac{C}{B^2} = \frac{N_C}{12 \pi^2} \frac{\beta + 2}{(2 \beta - 1) (3 - \beta)} \hk .
\ee
In view of eqs. (\ref{X10}) and (\ref{X14}) we have the following relations
between the infrared exponents
\be
\label{X16}
\alpha = \gamma = 2 \beta - 1\hk ,
\ee
so that $\omega (k)$ and $\chi (k)$ behave in exactly the same way in the
infrared. Combining eq. (\ref{X11}) and (\ref{X15}) we can eliminate the
constant $B$ and obtain 
\be
\label{X17}
\frac{A}{C} = \frac{(2  \beta - 1)(3 - \beta)}{\beta (\beta  + 1)} \hk .
\ee
\bi

\no
Unfortunately the gap equation (\ref{17-4}) cannot be treated in the same fashion.
This is because the integrand in $I_\omega (k)$ (\ref{17-5}) 
contains explicitly the external
momentum $k$. However, one can show, without resorting to the
angular approximation, that in the infrared
$k \to 0$ the quantities $\omega (k)$ and $\chi (k)$ approach each other, i.e.
 $\omega (k)$ and $\chi (k)$ do not only have the same divergent infrared 
behaviour
$(\alpha = \gamma)$ as shown above, but also have the same infrared strength
\be
\label{X18}
A = C \hk .
\ee
For this purpose consider the full gap equation (\ref{17-4}) 
in the limit $k \to 0$
assuming for the moment, that the ultraviolet diverging integrals have been
regularized. Renormalizaton carried out in the next section will remove the
divergent constant $I^0_\omega$ (\ref{17-5a}) and will introduce finite
renormalization constants which, however, become irrelevant in the infrared
limit $k \to 0$ compared to the diverging quantitites $\omega (k)$ and $\chi
(k)$.
Hence, ignoring infrared finite terms, the gap equation (\ref{17-4}) becomes
\be
\label{28-x1}
\omega^2 (k \to 0) = \chi^2 (k \to 0) + I_\omega (k \to 0) \hk .
\ee
Since the integral $I_{\omega} (k)$ (\ref{17-5}) is ultraviolet divergent the dominant
contributions to this integral must come from the large momentum region. For large 
but finite $q$ and $k \to 0$ in the integrant of $I_{\omega} (k)$ (\ref{17-5}) we can 
omit $\chi (q)$ and $\omega (q)$ compared to the divergent quantities $\omega (k \to 0)$, 
$\chi (k \to 0)$ yielding
\be
\label{28-x2}
I_\omega (k \to 0) = \left[ \chi^2 (k) - \omega^2 (k) \right] \hs \hs \hs \hs \hs \hs \hs \nonumber \\
\cdot \frac{N_C}{4} \int
\frac{d^3 q}{(2 \pi)^3} \frac{d^2 ({\bf k} - {\bf q}) f ({\bf k} - {\bf q})}
{ \lk {\bf k} - {\bf q} \rk^2 \omega (q)} \hs 
\ee
is ultraviolet finite. Indeed with the above found ultraviolet behaviour
(\ref{G17}) we obtain
\begin{align}
\int d q \frac{d^2 (q) f (q)}{\omega (q)} \sim \int d q \frac{1}{q (\ln
q /\mu)^{\frac {3}{2}} } = - 2 \int d q \frac{d}{d q} \frac{1}{(\ln q /\mu)^{\frac {1}{2}}} \hk .
\end{align}
With the singular behaviour of $\omega (k)$ and $\chi (k)$ for $k \to 0$ eqs.
(\ref{28-x1}), (\ref{28-x2}) obviously imply
\be
\chi (k \to 0) = \omega (k \to 0) \hk .
\ee
The same relation is also found in the full numerical solution of 
the renormalized gap
equation given in section \ref{sec7}. \\
%
With the relation (\ref{X18}) it follows from eq. (\ref{X17}) $\beta = {1}$ and
in view of eq. (\ref{X16}) $\alpha = \gamma = 1$. Thus, we obtain the following
infrared behaviour 
\be
\label{X19}
\omega (k) = \chi (k) = \frac{A}{k} \hk , \hk d (k) = \sqrt{\frac{8 \pi^2
A}{N_C}} \cdot \frac{1}{k}
 \hk .
\ee
Finally, we note, that the above obtained infrared behaviour for the ghost
propagator $d (k) \sim \frac{1}{{k}}$ is precisely the one, which is needed
to produce a linear rising confinement potential from the Coulomb energy, when
one uses the leading infrared
 approximation $f (k \to 0) = 1$ for the Coulomb form factor.
\bi

\no
\section{Renormalization \label{sec6}}
\bi

\no
The integrals occuring in the Schwinger-Dyson equations are divergent and
require regularization and renormalization. The renormalization of the
Schwinger-Dyson equations in Coulomb gauge has been already discussed in ref.
\cite{R16}. However, our equations differ from those of
ref. \cite{R16} due to the presence of the curvature $\chi$. The latter will
introduce new features, which require separate discussions. In ref. \cite{R17}
the curvature was also included but not fully: In the gap equation the curvature
was omitted under the loop integral in the Coulomb term, but it is precisely this dependence on the
curvature, which gives rise to the new troublesome features.
\bi

\no
For simplicity, we will use a 3-momentum cutoff $\Lambda$ as
ultraviolet regulator. We are aware of the fact, that such a procedure violates
gauge invariance and may give rise to spurious divergencies. Furthermore, the
approximate evaluation of the expectation value of the Coulomb term (neglecting
two-loop terms) will also introduce spurious ultraviolet divergent terms (see
below). The crucial point, however, is that neither the infrared nor the
ultraviolet behaviour of the quantities under interest (determined by the above
coupled Schwinger-Dyson equations) will depend on the specific regularization
and renormalization procedure used as will be demonstrated later. 
\bi

\no
After regularization the integrals $I_d (k), I_\chi (k), I_\omega (k)$ in the
Schwinger-Dyson equations become cutoff dependent $I_d (k) \to I_d (k, \Lambda)$
etc. . Consider first the equation (\ref{63}) 
for the ghost form factor
\be
\label{XX}
\frac{1}{d (k, \Lambda)} = \frac{1}{g} - I_d (k, \Lambda) \hk ,
\ee
where we have indicated, that after regularization the ghost form factor also
becomes cutoff dependent.
However, to assign a physical meaning to $d (k, \Lambda)$ we have
 to keep $d (k, \Lambda)$ independent of the cutoff
$\Lambda$ \footnote{As will be shown later (see eq. (\ref{132}), below), in the approximation
$f(k) = 1$ the static quark potential is given by $V = \frac{g^2}{2} G ( - \partial^2 ) G 
= \frac{g^2}{2} \frac{d}{g} \frac{1}{- \partial^2} \frac{d}{g} = \frac{1}{2} d ( - \partial^2 ) d $. 
Since quark potential is a renormalization group invariant we have obviously also to require
$d(k)$ to be a RG-invariant.} . 
This is achieved in the standard fashion by letting the coupling constant $g$ run with
the cutoff $\Lambda$. 
Independence of the ghost form factor $d (k, \Lambda)$ of
$\Lambda$ requires in view of (\ref{XX})
\be
\label{Z1}
\frac{d g (\Lambda)}{d \Lambda} = - g^2 (\Lambda) \frac{d I_d (k,
\Lambda)}{d \Lambda} \hk .
\ee
\bi

\no
For $\Lambda \to \infty$ we can ignore the $\Lambda$ dependence of $\omega (q)$
and $d (q)$ in the integrand of $I_d$ (\ref{63}) and find
\be
\label{G8}
\left. \frac{d I_d (k, \Lambda)}{d \Lambda} \right|_{\Lambda \to \infty} = 
\frac{N_C}{3}
\frac{1}{2 \pi^2} \frac{d (k = \Lambda, \Lambda)}{\omega (k = \Lambda, \Lambda)}
 \hk ,
\ee
which is independent of $k$. Since this quantity is positive $(\omega (k)$ has
to be strictly positive definite and $d (k)$ is positive inside the first Gribov
horizon) the solution $g(\Lambda)$ to eq. (\ref{Z1}) vanishes asymptotically for $\Lambda \to
\infty$. Furthermore, for $\Lambda \to \infty$ the integral $I_d (k = \Lambda,
\Lambda)$ (\ref{63}) 
remains finite (This is also explicitly seen from the angular
approximation (\ref{XZ1}).) Therefore from eq. (\ref{XX}) follows, that $d (k =
\Lambda, \Lambda)$ approaches asymptotically $g (\Lambda)$. 
Since furthermore $\omega (k \to \infty) \sim k$ we obtain
\be
\label{122}
\Lambda \frac{d g }{d \Lambda} = - {\beta} g^3 \hk ,
\hk \beta = \frac{{\beta}_0}{(4 \pi)^2} \hk ,  \hk {\beta}_0 = \frac{8 N_C}{3} \hk .
\ee
This result was also obtained in refs. \cite{R16}, \cite{R17}, 
which is not surprising, since the
 curvature does not enter
the equation for the ghost form factor. As discussed in ref. \cite{R16}
this coefficient should not be compared with the canonical perturbative 
expression
of $\beta_0 = 11 N_C / 3$. In the present approach the running coupling constant
can be extracted from the Coulomb term. One finds then (ref. \cite{R16})
$\beta_0 = 12 N_C /3$ instead of  $\beta_0 = 11 N_C / 3$. The difference is due
to the absence of the perturbative contribution due to the emission and
absorption of transverse gluons, when taking the expectation value of the
Hamiltonian. 
\bi

\no
Eq. (\ref{122}) has the well-known solution
\be
g^2 (\Lambda) = \frac{g^2 (\mu)}{1 + \beta g^2 (\mu) \ln \lk \frac{\Lambda^2}{\mu^2} \rk} 
\hk ,
\ee
which shows that for $\Lambda \to \infty$ the asymptotic behavour of $g
(\Lambda)$ is
\be
\label{147}
g^2 (\Lambda) = \frac{1}{\beta \ln \lk \frac{\Lambda^2}{\mu^2} \rk} \hk ,
\ee
in accordance with asymptotic freedom. The same behaviour was obtained in eq.
(\ref{19-9}) for the ghost form factor $d (k = \Lambda \to \infty)$. This shows,
that, indeed, asymptotically the ghost form factor approaches the running coupling
constant
\be
d (k = \Lambda \to \infty) \to g (\Lambda) \hk .
\ee
The integral
$I_d (k, \Lambda)$ (\ref{63}) 
is logarithmically ultraviolet divergent (This is also explicitly seen in the
angular approximation (\ref{XZ1}).) The equation (\ref{XX}) can
be renormalized by subtracting the same equation at an arbitrary renormalization
scale $\mu$
\be
\label{G5}
\frac{1}{d (\mu, \Lambda)} = \frac{1}{g (\Lambda)} - I_d (\mu, \Lambda) 
\ee
yielding
\be
\label{G6}
\frac{1}{d (k, \Lambda)} = \frac{1}{d (\mu, \Lambda)} - \left[ I_d (k, \Lambda) 
- I_d (\mu, \Lambda) \right] \hk .
\ee
Eq. (\ref{G5}) shows how (for a fixed renormalization scale $\mu$) the coupling
constant $g (\mu, \Lambda)$ has to run with $\Lambda$ in order, that the ghost
form factor $d (k)$ becomes independent of the cutoff. For $\Lambda \to \infty$
this dependence $g (\Lambda)$ is given by eq. (\ref{147}).
\bi

\no
The difference $I_d (k, \Lambda) - I_d (\mu, \Lambda)$ is ultraviolet finite, so
that we can take the limit $\Lambda \to \infty$ in eq. (\ref{G6})
\be
\label{nr1}
\frac{1}{d (k)} = \frac{1}{d (\mu)} - \lim\limits_{\Lambda \to \infty} \left[ I_d
(k, \Lambda) - I_d (\mu, \Lambda ) \right] \hk ,
\ee
where we have put $d (k, \Lambda \to \infty) = d (k)$ etc. This is the desired
finite Schwinger-Dyson equation for the ghost form factor, which contains the
arbitrary renomalization constant $d (\mu)$. 
\bi

\no
We use the same minimal subtraction procedure to renormalize the equation for
the curvature (\ref{71}) and the Coulomb form factor (\ref{13-XXX}). 
Subtracting the equation (\ref{71}) and (\ref{13-XXX}) once at the renormalization
scale $\mu$ yields
\be
\label{G10}
\chi (k) = \chi (\mu) + \Delta I_\chi (k) \hk \\
\label{nr4}
f(k) = f(\mu) + \Delta I_f (k) \hk ,
\ee
where the difference
\be
\Delta I_\chi (k) = I_\chi (k,  \Lambda) - I_\chi (\mu, \Lambda) \\
\Delta I_f (k) = I_f (k,  \Lambda) - I_f (\mu, \Lambda)
\ee
is ultraviolet finite, so that the limit $\Lambda \to \infty$ can be taken. The
finite quantities $\chi (\mu)$ and $f(\mu)$ are new renormalization constants, on which our
solutions, in principle, will depend. However, we will find later, that
this dependence is very mild and that neither the infrared nor the ultraviolet
behaviour of our solutions will depend on $\chi (\mu)$ and $f(\mu)$. \\
%
The renormalization of the gap equation (\ref{17-4}) 
is more involved, when the curvature
term is included. To
renormalize the gap equation we first follow the minimal subtraction procedure
(applied above to the ghost form factor and to the curvature). This removes the
(quadratically) divergent constant $I^0_\omega$ (\ref{17-5a}) resulting in 
\be
\label{G11}
\omega (k)^2 &=& \omega (\mu)^2 + k^2 - \mu^2 + \chi (k)^2 - \chi (\mu)^2 \nonumber\\
& &+ I_\omega (k) - I_\omega (\mu) \hk .
\ee
\\

Unfortunately, the resulting expression $I_\omega (k) - I_\omega (\mu)$ is still
diverging. This is a consequence of the one-loop approximation, used 
when calculating
the expectation value of the Coulomb term (resulting in the factorization
(\ref{x5})) 
and also when taking the variation
(functional derivative) of this term with respect to $\omega (k)$ to obtain the
gap equation. As is well known the gauge invariance is not maintained in the loop
expansion order by order \cite{R20}. As a consequence, truncating
the loop expansion at a given order results in spurious divergencies, which are
cancelled by higher order terms. Such spurious divergent terms should therefore
be omitted. To identify the spurious (divergent) 
terms in the gap equation, we rewrite the
diverging integral in the form
\be
\label{G12}
I_\omega (k, \Lambda) = I^{(2)}_\omega (k, \Lambda) + 2 
\chi (k) I^{(1)}_\omega (k, \Lambda) \hk ,
\ee
where
\begin{widetext}
\be
\label{G13}
I^{(n)}_\omega (k, \Lambda) & = & \frac{N_C}{4} \int^{\Lambda} \frac{d^3 q}{(2 \pi)^3} \lk 1 
+ (\hat{{\bf k}} \hat{{\bf q}})^2 \rk  
\cdot \frac{d ({\bf k} -{\bf q})^2 f ({\bf k} - {\bf q})}{({\bf k} - {\bf q})^2} 
\cdot \frac{\left[\omega ({\bf q}) - \chi ({\bf q})\right]^n - \left[\omega ({\bf k}) 
- \chi ({\bf k})\right]^n }
{\omega ({\bf q})} \hk .
\ee
\end{widetext}
The integral $I^{(2)}_\omega (k) $ is quadratically divergent. Its divergent part
is, however, independent of the external momentum $k$, so that one subtraction
eliminates this divergence, i.e. $I^{(2)}_\omega (k) - I^{(2)}_\omega (\mu)$ is
finite. The troublesome term in eq. (\ref{G12}) is the second one, which is
linearly divergent. Due to the
momentum dependent factor $\chi (k)$ one substraction does not eliminate the
divergency
\begin{widetext}
\be
\label{G15}
I_\omega (k, \Lambda) - I_\omega (\mu, \Lambda) & = & 
\underbrace{\left[I^{(2)}_\omega (k, \Lambda) - I^{(2)}_\omega
(\mu, \Lambda)\right]}_{\mbox{finite}} 
+ 2 \chi (k) \underbrace{\left[ I^{(1)}_\omega (k, \Lambda) - I^{(1)}_\omega (\mu,
\Lambda)
\right]}_{\mbox{finite}}  + 2 \left[ \chi (k) - \chi (\mu) \right] \underbrace{I^{(1)}_\omega
(\mu, \Lambda)}_{\mbox{divergent}} \hk , \hk \hk
\ee
\end{widetext}
while $I^{(1)}_\omega (k) - I^{(1)}_\omega (\mu) $ is finite, 
the last term is still diverging. Note, that this term disappears, when the
curvature is ignored $(\chi (k) \to 0)$ as done in ref. \cite{R16}\footnote{In
ref. \cite{R17} the curvature was included but was omitted under the integrals,
so that this troublesome term does not appear.}. 
(However, we will later find that it is of crucial importance to fully
keep the curvature.)
Due to its $k$ dependence a further substraction would not eliminate this
 divergency.
This is the type of spurious term discussed above, whose singular part
 violates gauge
invariance and would be cancelled by higher order loop terms. Once these higher
order loop terms are included the divergencies are cancelled and what is left
from $I^{(1)}_\omega (\mu, \Lambda)$ is
a finite contribution, which we denote by $I'^{(1)}_\omega (\mu)$. In the
following we will only keep this finite part $I'^{(1)}_\omega (\mu)$, which
unfortunately is not explicitly known. However, we will later show, that this
unknown constant $I'^{(1)}_\omega (\mu)$ does not essentially influence the
solutions of the renormalized coupled Schwinger-Dyson equations. \\
%
Applying the above described renormalization procedure (i.e. one
subtraction and replacing $I^{(1)}_\omega (\mu, \Lambda)$ by its finite part
$I'^{(1)}_\omega (\mu)$) the gap equation (\ref{G11}) becomes
\begin{widetext}
\be
\omega (k)^2 & = & k^2 + \chi (k)^2 + \omega (\mu)^2 - \mu^2 - \chi (\mu)^2
 + \left[ I^{(2)}_\omega (k, \Lambda) - I^{(2)}_\omega (\mu, \Lambda) \right] \nonumber\\
& & + 2 \chi (k) \left[
I^{(1)}_\omega (k, \Lambda) - I^{(1)}_\omega (\mu, \Lambda) \right]
+ 2 \left[ \chi (k) - \chi (\mu) \right] I'^{(1)}_\omega (\mu) \hk .
\ee
Inserting here for $\chi (k)$ its renormalized value (\ref{G10}) we obtain
\be
\label{nr2}
\omega (k)^2 & = & k^2 - \mu^2 + \Delta I_\chi (k)^2 + \xi \Delta I_\chi (k) 
+ \left[ I^{(2)}_\omega (k, \Lambda) - I^{(2)}_\omega (\mu, \Lambda) \right] \nonumber\\
& & + 2 \left[ \chi (\mu) + \Delta I_\chi (k) \right] \left[ I^{(1)}_\omega (k, \Lambda) 
- I^{(1)}_\omega (\mu,\Lambda) \right] +
\omega (\mu)^2  \hk .
\ee
\end{widetext}
All together there are five 
renormalization constants $d (\mu), \chi (\mu), f(\mu), \omega
(\mu)$ and $\xi = 2 [ \chi (\mu) + I'^{(1)}_\omega (\mu) ]$. Later 
one we shall demonstrate that the
self-consistent solution to the coupled Schwinger-Dyson equations does not
sensitively depend on the detailed values of these renormalization constants
except for $d (\mu)$.  \\
%
From our previous discussions it should be clear, that  the ultraviolet  
behaviour of the self-consistent solution 
does not at all depend on these
renormalization constants. We will later also show, that the infrared behaviour
does not depend on the precise value of the renormalization constants $\chi
(\mu), f (\mu), \omega (\mu)$ and $\xi$, while the ghost form factor
$d (k)$ depends crucially on $d (\mu)$ and only for one particular (critical) value
of $d (\mu)$ (for which $1 / d (\mu \to 0) \to 0$) 
the coupled Schwinger-Dyson equations will
have a solution, consistent with the infrared behaviour found in section \ref{sec5b}.
\bi

\no
\section{Numerical results \label{sec7}}
\bi

\begin{figure}
\includegraphics[scale=0.43,bb=24 32 600 436,clip=]{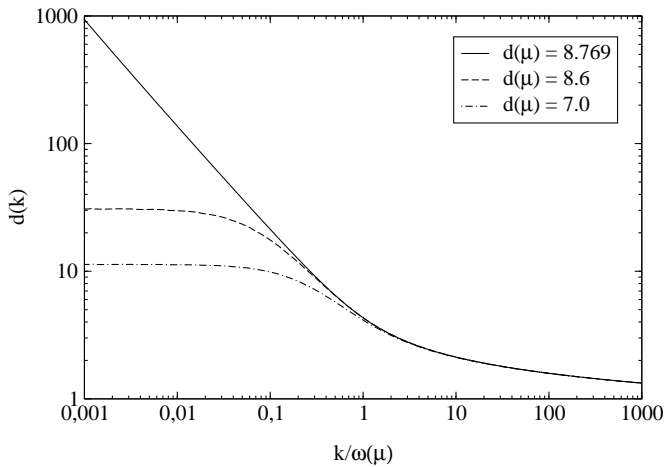}
\caption{\label{fig1} Solution for the ghost form function $d (k)$ for different
renormalization constants $d (\mu) = 7.0, 8.6$ and $8.716$.}
\end{figure}

\begin{figure}
\includegraphics[scale=0.43,bb=47 33 602 436,clip=]{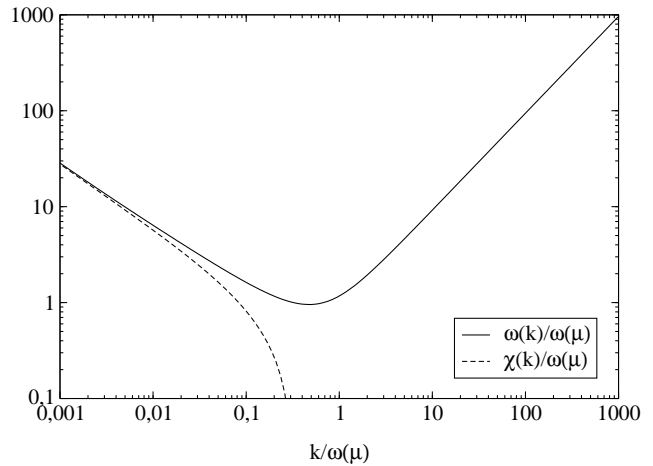}
\caption{\label{fig2} Solution for the gap function $\bar{\omega} (k) $ for $\bar{\xi} = 0$  }
\end{figure}

\begin{figure}
\includegraphics[scale=0.43,bb=47 33 602 436,clip=]{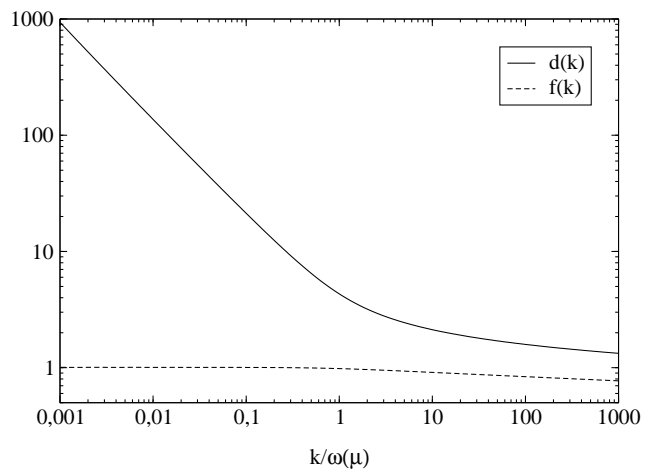}
\caption{\label{fig3} Ghost form function $d (k)$ and 
Coulomb correction $f (k)$ for $\bar{\xi} = 0$}
\end{figure}

\no
In this section we present the results of the numerical solutions to the coupled
Schwinger-Dyson equations (\ref{nr1}), (\ref{nr2}), (\ref{G10}), (\ref{nr4})
for the ghost form factor $d (k)$, the gluon energy
$\omega (k)$, the curvature $\chi (k)$ and 
the Coulomb form factor $f (k)$. For this purpose it is convenient to introduce
dimensionless quantities. We will rescale all dimensionfull quantities with
appropriate powers of the gluon energy $\delta := \omega (\mu)$ at an arbitrary renormalization point
$\mu$.
The rescaled dimensionless quantities will be indicated by a bar
\be
\label{GG2}
\bar{k}  =  \frac{k}{\delta} \hk , \hk \bar{\omega} (\bar{k}) = \frac{\omega
(k = \bar{k} \delta)}{\delta} \hk , \hk
\bar{\chi} (\bar{k})  =  \frac{\chi
(k = \bar{k} \delta)}{\delta} \hk .
\ee
The form factors $d (k)$ and $f (k)$ are dimensionless. \\
%
Before solving the coupled Schwinger-Dyson equations we have to fix the
renormalization constants
\be
\label{GG3}
d (\mu) \hk , \hk \bar{\xi} = 2 [ \bar{\chi} (\mu) + \bar{I}'^{(1)}_\omega (\mu) ] \hk ,
 \hk \bar{\chi} (\mu) \hk, f (\mu) \hk .
\ee
Note, that $\omega (\mu)$ has been absorbed into the dimensionless quantities. \\
%
The coupled Schwinger-Dyson equations are solved by iteration without resorting
to the angular approximation. 
To carry out the calculation consistently to one loop order, we should use the
leading expression for the Coulomb form factor $f(k)=1$ (see eq. (\ref{13-XXX})).
However, then we loose the anomalous dimension of $f(k) (\sim 1/\sqrt{\ln k^2/\mu^2},k \to \infty)$ 
which is needed for the
convergence of certain loop integrals. To keep the anomalous dimensions of $f(k)$
and at the same time include as little as possible corrections to the leading order
$f(k)=1$ we replace the ghost form factor $d(k)$ in the equation for $f(k)$ by its bare
value $d(k)=1$ and use $f (\mu) = 1$ . \\
The integrals
were calculated by using the Gauss-Legendre method. In order to obtain an
accurate mapping of the infrared region, a logarithmical distribution of the
supporting points was used. The self-consistent solutions are shown in figures \ref{fig1}, 
\ref{fig2} and \ref{fig3} for the choice of the renormalization 
constants $\hk \bar{\xi} = 0 , \bar{\chi} (\mu) = 0$. 

The value of the remaining renormalization constant
$d (\mu)$ has been specified as follows: \\

%
Consider the equation for the ghost form factor. The curvature $\bar{\chi}  (k)$, the
Coulomb form factor $f (k)$ as well as the renormalization constants $\bar{\xi}$ and
$\bar{\chi} (\mu)$ do not enter this equation. Thus, for given $\bar{\omega} (k)$ the
solution $d (k)$ depends only on the renormalization constant
$d (\mu)$. Figure  \ref{fig1} shows the solution to the ghost form factor for
various values of the renormalization constant $d (\mu)$ keeping
$\bar{\omega} (k)$ fixed to the solution shown 
in figure  \ref{fig2}. It is seen, that
all solutions have the same ultraviolet behaviour independent of the
renormalization constant $d (k)$. Furthermore, this ultraviolet
behaviour is consistent with the asymptotic solutions (\ref{19-9}) 
found in section \ref{sec5} 
(Note, that a double logarithmic plot 
is used)! The infrared behaviour of $d(k)$ depends, however, on the actual value of $d (\mu)$. 
For $d (\mu)$ smaller than some critical value $d_{cr}$ the curves approach a constant for $k
\to 0$. At a critical $d (\mu) = d_{cr}$ the ghost form factor
diverges for $k\to 0$ and above the critical value $d > d_{cr}$ no solution to the ghost form
factor exists. We have adopted the critical value $d (\mu) = d_{cr}$
as the physical value for the following reasons:
\begin{itemize}
\item[i)] In $D = 3$ (which will be considered elsewhere) a self-consistent 
solution to the coupled Schwinger-Dyson equations exists only for this critical value.
\item[ii)] Only the critical value produces an infrared diverging ghost form factor.
\item[iii)] The diverging ghost form factor is in agreement with our analytic 
studies
of the infrared limit of the Schwinger-Dyson equations considered in section \ref{sec5}
using the angular approximation (see eqs. (\ref{X19})).
\item[iv)] The divergent ghost form factor gives rise to a linear rising confining
potential as will be shown later on.
\end{itemize}

\begin{figure}
\includegraphics[scale=0.43,bb=24 33 602 436,clip=]{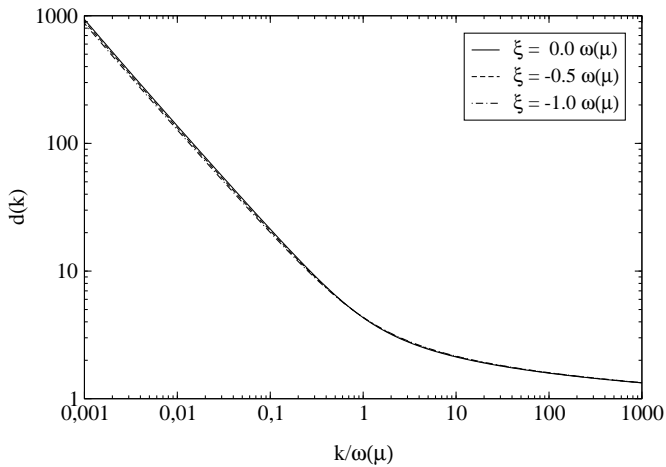}
\caption{\label{fig4} Ghost form function $d (k)$ for $\bar{\xi}= 0, - 0.5$ and $ - 1.0$ .}
\end{figure}

\begin{figure}
\includegraphics[scale=0.43,bb=23 33 602 436,clip=]{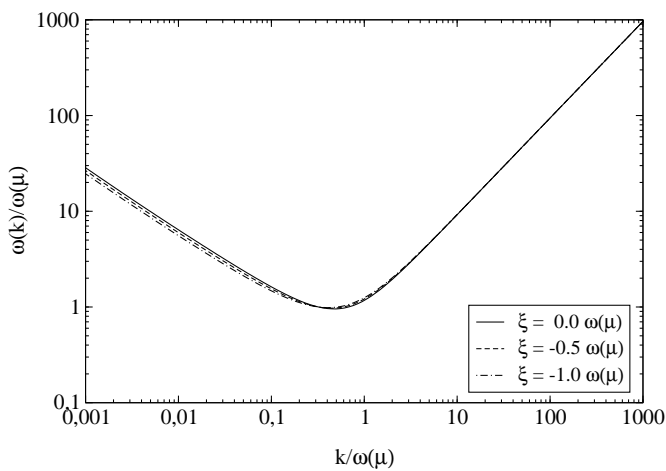}
\caption{\label{fig5} Gap function $\bar{\omega} (k)$ for $\bar{\xi}= 0, - 0.5$ and $ - 1.0$ . }
\end{figure}

\begin{figure}
\includegraphics[scale=0.43,bb=61 34 582 437,clip=]{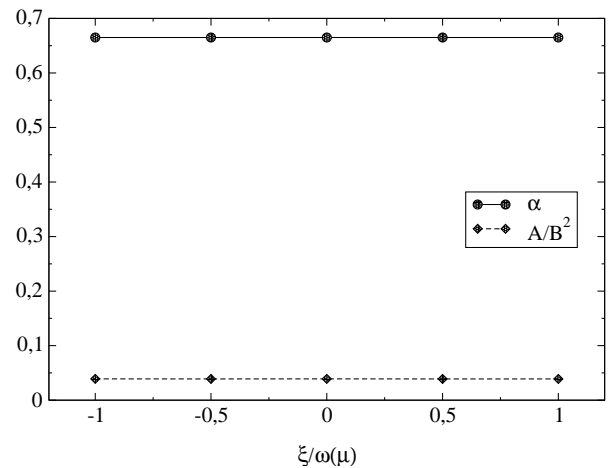}
\caption{\label{fig6} $A/B^2$ and $\alpha$ in dependence of $\bar{\xi}$  }
\end{figure}

\begin{figure}
\includegraphics[scale=0.43,bb=23 32 602 436,clip=]{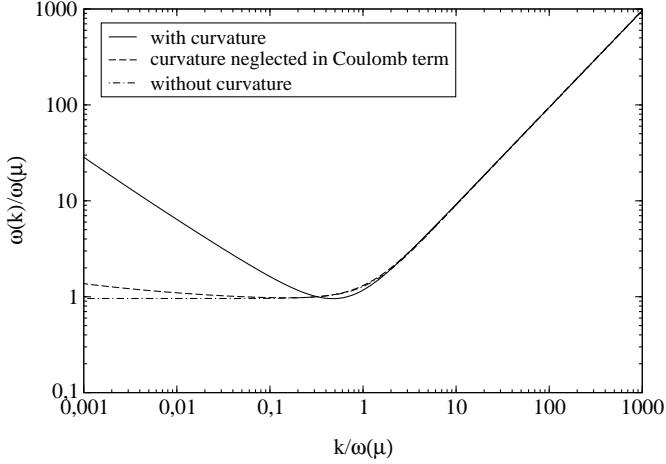}
\caption{\label{fig10f} Consistent solution of the gap function $\bar{\omega} (k)$
for different treatments of the curvature for $\bar{\xi} = 0$ .}
\end{figure}

\begin{figure}
\includegraphics[scale=0.43,bb=24 32 599 434,clip=]{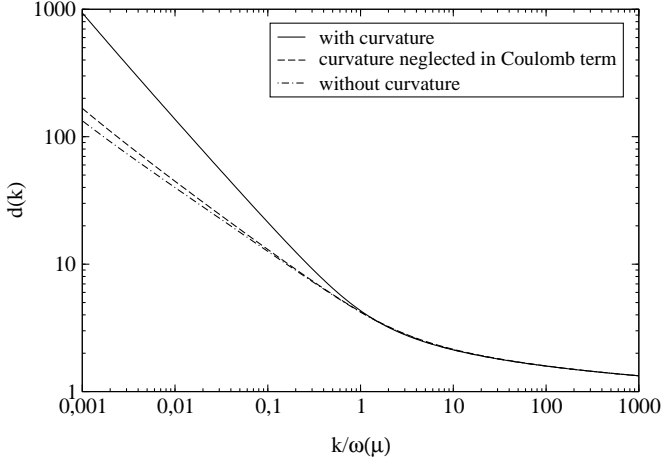}
\caption{\label{fig10g} Consistent solution of the ghost form function $d (k)$
for different treatments of the curvature for $\bar{\xi} = 0$ . }
\end{figure}

%
The critical $d (\mu) = d_{cr}$ is defined by $d^{-1} (k \to 0) \to 0$ which is referred to as
``horizon condition'' \cite{R19}. 
At the arbitrarily chosen (dimensionless) renormalization point $\bar{\mu} =
0.32$ the critical renormalization constant is given by $d (\bar{\mu}) =
8.716$. In all self-consistent solutions presented in this paper we have adopted
this critical value. \\
%
We have also investigated the dependence of the self-consistent solutions on the
remaining renormalization constants $\hk \bar{\xi}$ and $\bar{\chi} (\mu)$ (recall, that $d, 
\bar{\omega}
(\bar{k})$ are independent of $\hk \bar{\xi}$ and $\bar{\chi} (\mu)$). We have found, that our
self-consistent solutions change by less then $0.01^0/_{00}$, when $\bar{\chi} (\mu)$ is
varied in the intervall $[-1, 1]$. Thus there is practically no dependence of
our results on $\bar{\chi} (\mu)$. We have therefore put $\bar{\chi} (\mu) = 0$ in all calculations. \\
%
Figures \ref{fig4} and \ref{fig5} show the self-consistent solution for $d (k)$ and 
$\bar{\omega}
(k)$ for $\bar{\xi} = 0, - 0.5, - 1.0$. Both quantities show only very slight
variations with $\bar{\xi}$ up to a (dimensionless) momentum of order one. The
ultraviolet behaviour is independent of $\bar{\xi}$ and in agreement with our
analytic results obtained in section \ref{sec5}. Furthermore, also the infrared
behaviour of $d (k)$ and $\bar{\omega}
(k)$ is independent of $\bar{\xi}$. \\
%
%
Our analysis of the infrared behaviour of the solutions to the Schwinger-Dyson
equations using the angular approximation given in section \ref{sec5} has revealed,
that in this approximation the critical exponents $\alpha$ and $\beta$
 and the ratio $A / B^2$ of the amplitudes of $\bar{\omega}
(k)$ and $d (k)$, see eq. (\ref{X5}), are independent of the
renomalization constants $\hk \bar{\xi}$, $\bar{\chi} (\mu)$. Our numerical solutions confirm this
result even without resorting to the angular approximation. Figure \ref{fig6} shows these
quantities $A / B^2$ and the infrared exponent (\ref{X5}) 
$\alpha$ as function of $\hk \bar{\xi}$. There is practically no
dependence. We will therefore put $\hk \bar{\xi} = 0$ in the further
 numerical calculations. \\
%
The obtained numerical results are all in qualitative agreement with our
previous analytic investigations. For large $k \to \infty$ the gluon energy
$\omega (k) \sim \sqrt{k^2}$ is that of a non-interacting boson and the curvature
in orbit space $\chi (k) /\omega (k) \sim \frac{1}{\sqrt{\ln k /\mu}}$ vanishes
asymptotically. This is in agreement with the expectations of asymptotic
freedom. For $k \to 0$ the gluon energy $\omega (k)$ diverges reflecting the
absence of free gluons in the infrared, which is a manifestation of confinement.
While the gluon propagator $\frac{1}{\omega (k)} \to 0$ for $k \to 0$ 
is suppressed in the
infrared the ghost propagator $d (k) /k^2$ diverges for $k \to 0$. It is
worthwile noticing, that the same behaviour of the gluon and ghost propagators is
obtained in covariant Schwinger-Dyson equations, derived from the functional
integral in Landau gauge \cite{R21}, \cite{RX}. 
From figure \ref{fig2} it is seen, that for $k \to 0 \hk \hk \hk \omega
(k)$ approaches $\chi (k)$. As shown analytically in section \ref{sec5} 
this is a generic feature of our gap equation and
reflects the non-trivial metric  of the space of gauge orbits 
(given by the Faddeev-Popov matrix). This non-trivial metric is crucial for the
infrared behaviour of the theory and in particular for the confinement. This can
be seen from figures \ref{fig10f}, \ref{fig10g} , where we present the self-consistent solutions for 
$\omega
(k)$ and $d (k)$, when the curvature of the gauge orbit space is neglected by
putting $\chi (k) = 0$ as done in \cite{R16} or when neglecting the curvature
in the Coulomb term as done in \cite{R17}.
In these cases the infrared behaviour of $\omega (k)$ is
drastically different from the previous case, although we have still chosen the horizon condition
$d^{-1} (k \to 0) = 0$ as renormalization condition ($d(k)$ is still infrared divergent as can be
read off from figure \ref{fig10g}). In particular notice that $\omega (k \to 0) = \textrm{const.}$
when the curvature $\chi$ is completely neglected, i.e. $\alpha = 0$ in eq. (\ref{X5}). From the two
sum rules (\ref{X16}) for the infrared critical exponents follows $\beta = \frac {1}{2}$ and $\gamma = 0$.
Thus with the horizon condition as renormalization the neglect of the curvature in the Schwinger-Dyson
equation yields $d(k) \sim  \frac {1}{\sqrt{k}}$, $\omega (k) = \textrm{const.}$ and
$\chi (k) = \textrm{const.}$ for $k \to 0$ .
\bi

\no
\section{The Coulomb potential \label{sec8}}
\bi

\no
The vacuum expectation value of the 
Coulomb term of the Yang-Mills Hamiltonian can be interpreted as interaction
potential between static color charge densities $\rho^a ({\bf x})$. The static
quark potential can therefore be extracted from this term by taking the vacuum
expectation value and assuming, that the
color charge density $\rho^a ({\bf x})$ describes two static infinitely heavy
color charges
\be
\label{130}
\rho^a (x) = g q^a_{(1)} \delta \lk {\bf x} - {\bf x}_{(1)} \rk + g q^a_{(2)}
\delta \lk {\bf x} - {\bf x}_{(2)} \rk
\ee
located at  
${\bf x}_{(1)}$ and ${\bf x}_{(2)}$ and separated a distance 
${\bf x}_{(1)} - {\bf x}_{(2)} = {\bf r} $ 
 apart. This yields
\be
\label{131}
E_C & = & E^{(1)}_C + E^{(2)}_C 
+ q^a_{(1)} V^{a b} \lk {\bf x}_{(1)}  , {\bf x}_{(2)} \rk q^b_{(2)} \hk ,
\ee
where $E^{(1, 2)}_C$ are the (divergent) self-energies of the two separate
static quarks and 
\be
\label{132}
V^{a b} \lk {\bf x}_{(1)}  , {\bf x}_{(2)} \rk = g^2 \langle \omega | F^{a b} \lk
{\bf x}_{(1)}  , {\bf x}_{(2)} \rk | \omega \rangle
\ee
is the static quark potential, with $F^{a b} \lk {\bf x}, {\bf x}' \rk$ being 
the Coulomb propagator defined by eq. (\ref{13}). In the above considered one-loop
approximation the potential is color diagonal $V^{ab}=\delta^{ab} V$ and, with the explicit form of the
Coulomb propagator, is given by
\be
\label{135}
V \lk {\bf r} \rk =  \int \frac{d^3 k}{(2 \pi)^3} \frac{ d (k)^2 f
(k)}{k^2} e^{i {\bf k} {\bf r}} =  \int \frac{d^3 k}{(2 \pi)^3} e^{i
{\bf k} {\bf r}} V (k)  . \hk \hk \hk \hk
\ee

\begin{figure}
\includegraphics[scale=0.43,bb=32 32 591 436,clip=]{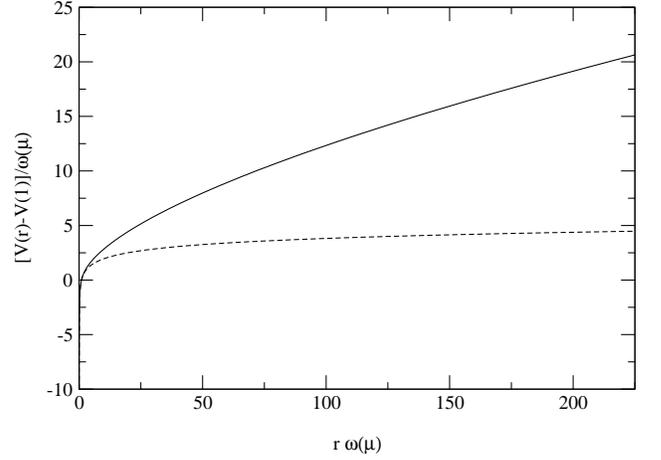}
\caption{\label{fig11} Coulomb Potential for $\bar{\xi} = 0$ with (full line) and without
inclusion of the curvature (dashed line). }
\end{figure}
Performing the integral over the polar angle, one finds
\be
\label{136}
V (r) = \frac{1}{2 \pi^2} \il^\infty_{0} d k { d (k)^2 f
(k)} \frac{\sin (k r)}{k r} \hk .
\ee
%
\no
Before presenting the numerical result for the Coulomb potential let us
consider its asymptotic behaviour for $k \to 0$ and $k \to \infty$. 
In section \ref{sec5b} we have found the infrared behaviour $f(k \to 0) = $ const. and
$d(k \to 0) \sim \frac {1}{k} $.
This yields precisely a linearly
rising Coulomb potential $V (k) \sim 1/k^4$.
Furthermore for $k \to
\infty$ the ghost form factor was found to behave as (see eq. (\ref{19-9}) $d
(k) \sim \frac{1}{\sqrt{\ln k /\mu}}$. 
Adopting the leading order expression for the Coulomb form factor
(see eq. (\ref{13-XXX}) and figure \ref{2a}) $f(k) = 1$ we find
\be
k^2 V (k) \sim \frac{1}{\ln k /\mu} \hk , \hk k \to \infty \hk .
\ee
This is precisely the behaviour found in ref. \cite{R25} in one-loop
perturbation theory. \\
The Coulomb potential calculated from the numerical solution to the coupled
Schwinger-Dyson equations is shown in figure \ref{fig11}. At small distance it is dominated
by an ordinary $\sim \frac{1}{r}$ potential, while at large distances it raises
almost linearly. The numerical analysis shows, that its Fourier transform
behaves for $k \to 0$ like $1/k^{3.7}$, while a strictly linearising potential
would require a $1/k^4$ dependence (In ref. \cite{R30} the power $1 / k^{3.6}$ was found).
When the curvature is neglected the gluon energy becomes infrared finite and
the Coulomb potential approaches a constant at $r \to \infty$. Thus both quark and
gluon confinement is lost when the curvature of the space of gauge orbits is discarded.
\bi

\no
\section{Summary and Conclusions \label{sec9}}
\bi

\no
In this paper we have solved the Yang-Mills Schr\"odinger equation for the
vacuum in Coulomb gauge by the variational principle using a trial wave function
for the Yang-Mills vacuum, which is strongly peaked at the Gribov horizon. Such
a wave functional is recommended by the fact, that the dominant infrared field
configurations lie on the Gribov horizon. Such field configurations include, in
particular, the center vortices, which have been identified as the confiner of
the theory. With this trial wave function the vacuum energy has been evaluated
to one-loop order. Minimization of the vacuum energy has led to a system of
coupled Schwinger-Dyson equations for the gluon energy, the ghost and Coulomb
form factor and for the curvature in orbit space. Using the angular
approximation these Schwinger-Dyson equations have been solved analytically in
both, the infrared and the ultraviolet regime. In the latter case, we have found
the familiar perturbative asymptotic behaviours. In the infrared the gluon
energy diverges indicating the absence of free gluons at low energies, which is
a manisfestation of confinement. The ghost form factor is infrared diverging and
gives rise to a linear rising static quark potential. The asymptotic analytic
solutions for both $k \to 0$ and $k \to \infty$ are reasonably well reproduced by the
full numerical solutions of the coupled Schwinger-Dyson equations. Our
investigations show, that the inclusion of the curvature, i.e. the proper
metric of orbit space, given by the Faddeev-Popov determinant is crucial in
order to obtain the confinement properties of the theory. When the curvature is
discarded (using a flat space of gauge connnections) free gluons exist even for
$k \to 0$ and the static quark potential is no longer confining. 
\bi

\no
The results obtained in the present paper are quite encouraging and call for
further studies. In a subsequent paper we will investigate along the same lines
the $2 + 1$-dimensional Yang-Mills theory, which (up to a Higgs field)
can be considered as the high
temperature limit of the $3 + 1$-dimensional theory. 
It would be also interesting to calculate the spatial Wilson loop in order to check
whether the relation $\sigma_{coul} \approx 3 \sigma $ found on the lattice \cite{R26}
is reproduced. Furthermore, the spatial t'Hoft loop should be calculated using the 
continuum representation derived in \cite{R29}.
Eventually one should
include dynamical quarks, since the ultimate goal should be the description of the
physical hadrons.
\bi

\no
\section*{Note added}
\bi

\no
After this work was completed we have been able to show that the infrared limit of the 
Yang-Mills wave functional (\ref{47xy}) is independent of the power $\alpha$ of the 
Faddeev-Popov determinant \cite{R100}.
\bi

\no
\section*{Acknowledgements}
\bi

\no
The authors are grateful to R. Alkofer, C.S. Fischer, J. Greensite, O. Schr\"oder, 
E. Swanson, A. Szczepaniak and D.
Zwanziger for useful
discussions. They also thank O. Schr\"oder for a critical reading of the manuscript and 
useful comments.
This work was supported by Deutsche Forschungs-Gemeinschaft under contract DFG-Re 
856.

\end{document}